\documentclass[preprint,journal]{vgtc}            

\onlineid{1865}

\vgtccategory{Research}

\vgtcpapertype{application/design study}

\usepackage{pifont}

\newcommand{\toolname}[0]{{DaedalusData\xspace}}

\newcommand{\videotimestamp}[1]{\textit{#1}}

\newcommand{\cmark}{\ding{108}}
\newcommand{\xmark}{\ding{109}}

\title{\toolname{}: Exploration, Knowledge Externalization and Labeling of Particles in Medical Manufacturing -- A Design Study}

\author{%
  \authororcid{Alexander Wyss}{0009-0009-2763-3186}
  \authororcid{Gabriela Morgenshtern}{0000-0003-4762-8797}
  \authororcid{Amanda Hirsch-Hüsler}{0000-0003-0090-2728}
  \authororcid{Jürgen Bernard}{0000-0001-8741-9709}
}

\authorfooter{
  \item
  	Alexander Wyss, University of Zurich \& Roche
  	E-mail: alexander.wyss@protonmail.com
   \item 
        Gabriela Morgenshtern, University of Zurich \& Digital Society Initiative Zurich
        E-mail: morgenshtern@ifi.uzh.ch
    \item 
        Amanda Hirsch-Hüsler, formerly Roche
        E-mail: a.hirschhuesler@gmail.com
    \item 
        Jürgen Bernard, University of Zurich \& Digital Society Initiative Zurich
        E-mail: bernard@ifi.uzh.ch
}

\abstract{
In medical diagnostics of both early disease detection and routine patient care, particle-based contamination of in-vitro diagnostics consumables poses a significant threat to patients.
Objective data-driven decision-making on the severity of contamination is key for reducing patient risk, while saving time and cost in quality assessment.
Our collaborators introduced us to their quality control process, including particle data acquisition through image recognition, feature extraction, and attributes reflecting the production context of particles.
Shortcomings in the current process are limitations in exploring thousands of images, data-driven decision making, and ineffective knowledge externalization.
Following the design study methodology, our contributions are a characterization of the problem space and requirements, the development and validation of \toolname{}, a comprehensive discussion of our study's learnings, and a generalizable framework for knowledge externalization.
\toolname{} is a visual analytics system that enables domain experts to explore particle contamination patterns, label particles in label alphabets, and externalize knowledge through semi-supervised label-informed data projections.
The results of our case study and user study show high usability of \toolname{} and its efficient support of experts in generating comprehensive overviews of thousands of particles, labeling of large quantities of particles, and externalizing knowledge to augment the dataset further.
Reflecting on our approach, we discuss insights on dataset augmentation via human knowledge externalization, and on the scalability and trade-offs that come with the adoption of this approach in practice.
}


\teaser{
  \centering
  \includegraphics[width=\linewidth, alt={\toolname{} Framework}]{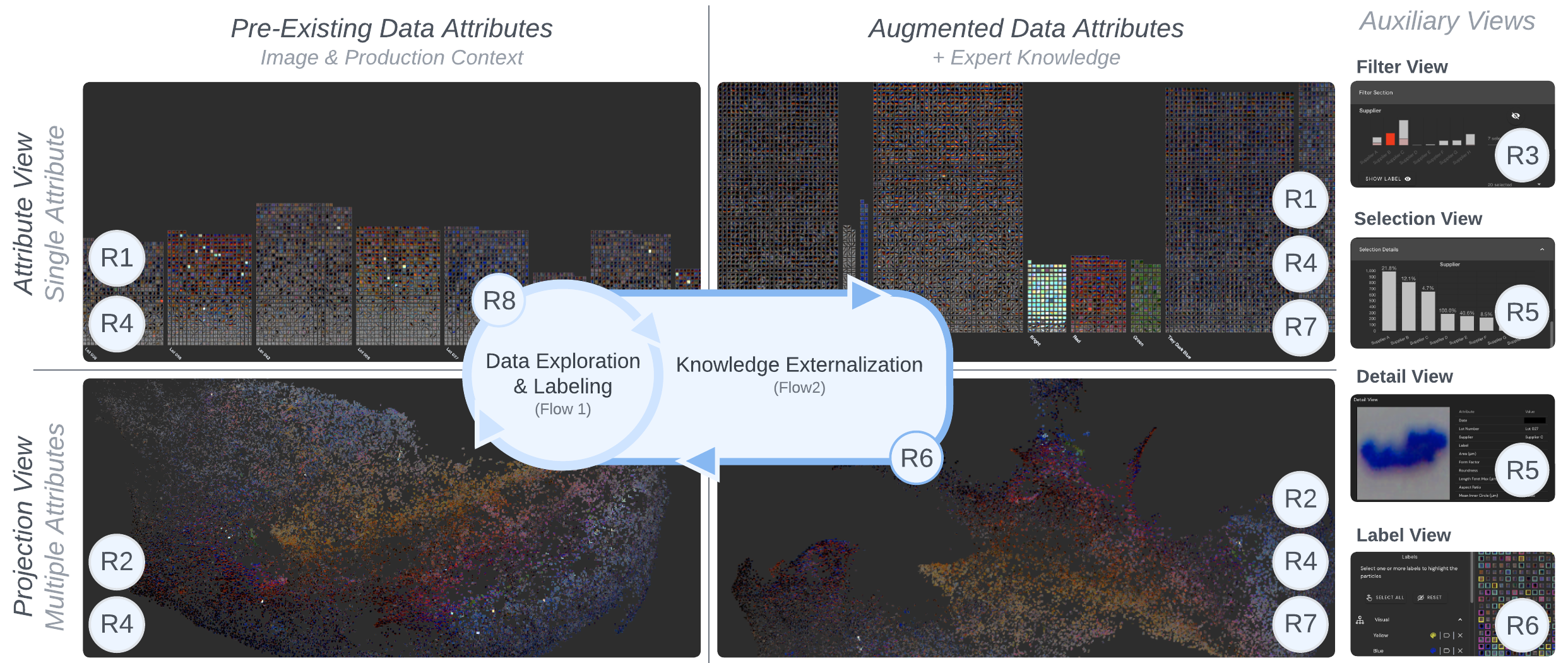}
  \caption{
    The \toolname{} framework supports two control modes for experts to steer the particle display with, shown here as a $2 \times 2$ matrix. 
    Vertical: Experts choose between the \emph{Attribute View} (for one attribute) and the \emph{Projection View} (for multiple user-specified attributes) to identify areas of interest, and discover similar particles to label. 
    Horizontal: Experts choose to explore either the \emph{Pre-Existing Data Attributes} (the Image \& Production Context), or to extend the exploration to \emph{Augmented Data Attributes} created through particle labeling (Expert Knowledge).
    This design study implements a systematic cross-cut of all four types of control, addressing expert-contributed design requirements (seen as ($R_n$) in the figure).
    \toolname{} enables experts to conduct two core workflows: \emph{Data Exploration and Labeling (Flow 1)}, and its extension, \emph{Knowledge Externalization (Flow 2)}. 
    On the right, we introduce four \emph{Auxiliary Views} that perform labeling, and facilitate interactive drill-downs, particle selection, and detailed analysis.
  }
  \label{fig:vis_concept}
}

\graphicspath{{figs/}{figures/}{pictures/}{images/}{./}} 

\usepackage{tabu}                      
\usepackage{booktabs}                  
\usepackage{lipsum}                    
\usepackage{mwe}                       
\usepackage{mathptmx}                  

\begin{document}


\firstsection{Introduction}

\maketitle

In modern medicine, the accuracy and reliability of In-Vitro Diagnostics (IVD) play a pivotal role in the timely detection and effective management of diseases. 
IVD refers to tests performed on samples, such as blood or tissue, taken from the human body.
These tests are essential for detecting diseases, monitoring health, and informing treatment decisions.
Therefore, the pursuit of uncontaminated diagnostic consumables (the materials used in IVD tests) is key to patient safety and healthcare excellence.
Diagnostic test contaminants, like foreign DNA / RNA in a Polymerase Chain Reaction (PCR) test, can lead to fatal errors in testing, causing false results or test abortion~\cite{YangSamuel.2004}.

To ensure consumable integrity, product quality engineers at Roche Diagnostics regularly review samples of a manufactured \emph{lot} (a batch of consumables produced in a given period), looking for signs of contamination by various particles.
Despite their expertise in assessing particle contamination, Roche's product quality engineers still face challenges in identifying patterns across these thousands of particles, augmenting the attributes of the particles' dataset with their experiential knowledge, and leveraging these for data-driven decision-making. 
This is vital for improving the diagnostic manufacturing process.
The lack of a system to standardize the examination of patterns at scale for the sample sources and material characteristics presents an opportunity to improve Roche's quality control processes.

To keep track of their observations of contamination, engineers take photographs of particles contaminating a sample, and store the image data along with production attributes: the supplier of the produced consumable, the production lot, and the production date.
However, visual characteristics of particle images can currently only be reviewed on an ad-hoc basis, due to dataset size and insufficient analytical means.
The number of particles found per consumable examined dictates whether the contents of a production lot can be used for diagnostics, or should be disposed of due to its particle contamination.
A more effective data-driven approach may better inform these decisions, reducing both costs and production waste, for ecological and economic benefit.

Using design study methodology, we collaborated with Roche product quality engineers to understand the problem space and the requirements of a solution. 
We then iteratively designed, developed, and validated \toolname{} as a solution.
\toolname{} is a visual analytics (VA) system that enables product quality engineers to explore particle data, discover new patterns and relationships between particles and their attributes, and externalize their tacit knowledge for data-driven decision-making in IVD particle contamination. 
\looseness = -1
It efficiently transforms expert knowledge into labels that can be rigorously and collaboratively analyzed for more informed decision-making.
We named our approach after the skilled and knowledgeable craftsman Daedalus from Greek mythology, highlighting the tool's focus on expert knowledge.

\toolname{} further allows experts to augment the underlying dataset with different context-sensitive label alphabets.
It supports views for interactive visualization of thousands of particles, designed to partition particles by a single attribute, or to project particles by dimensionality reduction using a user-selectable set of attributes.
For interactive data labeling, \toolname{} enables experts to drill down into the particle search space through interactive zooming, panning, filtering, and selection interactions. 
Auxiliary Views provide visual cues for filtering and selection-detail analysis.
\looseness = -1
\toolname{} supports knowledge externalization by facilitating the persistence, development, and extension of parallel \emph{label alphabets} for collaborative use.
The combined projection of particles based on both attributes \emph{and} labels, using a semi-supervised dimensionality reduction technique, provides visual cues to particle distributions and reveals yet-unlabeled particles.
By integrating expert knowledge directly into the visualization process, \toolname{} offers a more nuanced data exploration, enabling experts to uncover patterns and relationships that might otherwise remain undiscovered.

We evaluated \toolname{} through a user study and two case studies with Roche quality engineers.
The results show that \toolname{} efficiently generates meaningful, comprehensive particle data overviews, assists experts in accurately and efficiently labeling large numbers of particles, augments the particle dataset with their externalized knowledge, and is useful for generating actionable insights.

To summarize, our contributions are as follows:  
\begin{itemize}[noitemsep,leftmargin=*]
    \item We characterize the medical diagnostics domain, the problem space, the particle dataset, and the requirements for systematic knowledge capture within this domain; for the first time
    (Section \ref{sec:process_and_abstractions}).
    \item We develop and validate \toolname{} in accordance to the requirements, and address the challenges of handling large volumes of particle data, systematically capturing expert knowledge, and facilitating efficient and accurate labeling (Section \ref{sec:proposed_solution}, \ref{sec:evaluation}).
    \item We share insights gained through design study on the usability and scalability of \toolname{}, dataset augmentation and feature ideation via knowledge externalization,
    and labeling methodologies that balance human precision and automation (Section~\ref{sec:discussion_and_reflection}).
    \item We propose a generalizable framework for enhancing data exploration with data augmentation.
    Experts can utilize externalized domain semantics as user-specified attributes for \textit{label-informed} projections, as a similarity-preserving data positioning. (Figure \ref{fig:vis_concept}).
\end{itemize}


\section{Background}
IVD consumables play a critical role for analyzing patient samples such as blood, urine or tissue. 
Conducted outside the human body, diagnostic testing is essential for early disease detection, monitoring, treatment decisions, and consequently: patient care~\cite{10.1371/journal.pone.0149856}. 
Particle contamination of IVD consumables, especially in applications using the highly sensitive PCR technique, can lead to misleading results such as test failure, or false positives/negatives. 
These can cause serious patient harm, as PCR analyses for the presence of pathogens or hereditary diseases by amplifying samples of DNA / RNA collected from a patient.
This laboratory procedure facilitates the detection of a given gene sequence, so when particle contamination occurs, contaminant DNA fragments introduced into the PCR reaction are amplified alongside the sequence of interest, potentially falsifying results~\cite{YangSamuel.2004}. 
In addition to DNA, other types of contaminants, such as metal particles, can also interfere with the PCR process~\cite{Kuffel.2021}. 
Given the significant impact on test results and patient care, identifying patterns in particle contamination is essential.
To diagnose diseases early and administer the correct treatment, test results must be accurate and reliable.
This requires that the specimen, test equipment, and auxiliary test materials function properly, and be free from defects that could adversely affect the test result. 
Our research focuses on the issue of particle contamination of IVD consumables that come into direct contact with test samples.


\section{Related Work}
In this section, we cover related work on data exploration, knowledge externalization and data augmentation, interactive data labeling, and design studies within VA related to our domain problem.

\subsection{Exploratory Analysis of High-Dimensional Image Data} 
Interactive exploration tools are essential for identifying patterns and anomalies in large image datasets.
Pixplot~\cite{Duhaime16} and Collection Space Navigator~\cite{DBLP:journals/corr/abs-2305-06809} tackle the challenge of allowing to explore large image datasets through raster and projection-based views, allowing experts to customize the visual representation.
Systems like Vitessce~\cite{keller2021vitessce}, LFPeers~\cite{cagSachdeva2023}, and TissUUmaps~\cite{tissuumap} use linked views to propagate information into a dimensionality reduction projection, similar to \toolname{}.
UCLA~\cite{fujiwara2021interactive} introduces observation-level interactions to explore relationships and the comparability of labeled data groups.
WizMap~\cite{DBLP:journals/corr/abs-2306-09328} ensure better interpretability of projections by labeling regions of interest, an explainability approach that \toolname{} enhances with attribute contextualizatios for selected images.
Beyond the rich image exploration support provided by many approaches, \toolname{} incorporates expert-generated knowledge attributes through label-informed projections, enhancing the discovery of image clusters of interest.

\subsection{Knowledge Externalization and Data Augmentation} 
Interactive knowledge externalization allows for data augmentation, essential for enabling analysis and decision-making downstream. 
Methods for eliciting tacit knowledge can be borrowed from cognitive science methodologies, like self-reports ~\cite{Schwarz1999} and cognitive task analysis ~\cite{chipman2000introduction,klein1989critical,militello1998applied}.
Both van Wijk ~\cite{wijk05} and Wang et al. ~\cite{wangJDLRC09} established foundational knowledge externalization methodologies within visualization and VA, emphasizing the need for direct manipulation~\cite{endertBN13} of interfaces when eliciting tacit knowledge from users.
Interfaces enabling knowledge-assisted interactions, such as annotations ~\cite{WagnerSHRZA19,wang2024} and rankings~\cite{mistelbauer2012smart,boukhelifa2013evolutionary,cagBarth2023}, enable the externalization of expert preferences. 
These interactions extend to sorting tagged data~\cite{wang2024,chung2015knowledge}, assessing attribute importance~\cite{fujiwara2021interactive}, and selecting image regions of interest~\cite{bouali2016vizassist}. 
Expert-driven feedback mechanisms for knowledge externalization are diverse and facilitate a broad range of ontology frameworks, expert ontologies, rules, and graphical annotations
~\cite{lohfink2021knowledge,2019_Moritz_et_al,2020_Nie_et_al,2018_Stitz_et_al,vahc2015}.
\toolname{} addresses the need for knowledge externalization through its semi-supervised \textit{label-informed projections}, facilitating data augmentation and projection adaption to better accommodate experts' understanding of the data in context.
In this way, \toolname{} also enables the exploration of large image collections based on the experts' image context and rich, context-specific augmented metadata, extending the the data attributes and features available prior to tool execution.

\subsection{Interactive Data Labeling} 
Efficient and interactive labeling is crucial for creating ground truth image data~\cite{DBLP:conf/chi/FruchardMCH23}. 
\toolname{} enables experts to contribute knowledge directly to the labeling process, through data augmentation~\cite{Interactive_Lung9394090, MorphoClusters20113060} in human-in-the-loop approaches for semi-supervised learning~\cite{sliceProp10403175, DBLP:journals/tac/ZhuKKST18}.
\toolname{} does not rely on classification and active learning models~\cite{Ren22, Zhang2022, BuddRK21,KumarG20}, based on 
a) the rationale to let humans externalize knowledge only for particles of interest, 
b) the awareness that human instance selection can compete with active learning~\cite{bernard_jurgen_comparing_2017}, and 
c) possible issues of active learning in different process phases like bootstrap problems and biased instance selections~\cite{attenberg2010, LUGHOFER2012,bernardCGF2018,euroVisShort2018}.
\toolname{} uses human-centered labeling methodologies with a high degree of human control, known as \emph{user-based active learning}~\cite{seifert2010}, \emph{interactive learning}~\cite{hoeferlin2012}, and \emph{Visual-Interactive Labeling (VIAL)}~\cite{bernard2018vial}.
Example applications include soccer player analysis~\cite{ivapp2017Soccer}, human pose clustering and classification~\cite{vda2017}, personalized music classification~\cite{euroVA2018music}, football player classification~\cite{pacificVAST2019}, topic modeling~\cite{felix2018exploratory}, and expert-validated predictive modeling in intensive care~\cite{morgenshtern2023riskfix}.
Finally, \toolname{} increases labeling efficiency through multi-particle selection, similar to Hoi et al.~\cite{hoi2006large} and Benato et al.~\cite{Benato2018}, who also combined multi-instance selection with dimensionality reduction. 

\subsection{Related Design Studies}
Our approach leverages collaborative, iterative development with experts to align with user needs~\cite{DBLP:series/lncs/Munzner08,DBLP:journals/tvcg/SedlmairMM12,DBLP:conf/beliv/Sedlmair16,DBLP:journals/tvcg/GarrisonMSOHB21,DBLP:journals/tvcg/SomarakisUKLH21,Interactive_Lung9394090}. 
In the medical manufacturing domain, design study approaches for IVD often focused on image segmentation, rather than exploration and labeling~\cite{DBLP:journals/tvcg/SomarakisUKLH21, Interactive_Lung9394090, DBLP:journals/tvcg/YuanRWG13}.
We, in contrast, make extensive use of dimensionality reduction and linked views, like design studies in high-dimensional data exploration~\cite{DBLP:journals/mva/HanFNLLN23,keller2021vitessce,DBLP:journals/tvcg/EirichBJSSFSB22}.
Further inspiration comes from approaches focusing on attribute-centered organizations and semantic data groupings~\cite{blumenschein2018smartexplore,wenskovitch2021examination}, which inform the attribute-based views of \toolname{}.
Design studies providing expert-user labeling interfaces include approaches enhancing the accuracy and efficiency of data annotation, crucial in domains requiring highly informed decision-making, like healthcare.
Strategies include crowd feedback elicitation~\cite{morgenshtern2023riskfix, park2024dynamiclabels}, dimensionality reduction projections~\cite{fujiwara2021interactive,wenskovitch2021examination,DBLP:journals/tvcg/EirichBJSSFSB22}, direct image labeling~\cite{chang2016alloy,chang2017revolt},
and direct manipulation of a visualization~\cite{morgenshtern2023riskfix, coscia2024deepsee}.
Finally, DynamicLabels~\cite{park2024dynamiclabels} reduces manual labeling by highlighting the potential of crowdsourced annotations to refine label sets, suggesting labels automatically.
Our design study integrates labeling into a high-dimensional data exploration and annotation workflow, for the decision-making support of experts in medical diagnostics manufacturing.

\section{Process and Abstractions}
\label{sec:process_and_abstractions}

\begin{figure*}[t]
    \centering
    \includegraphics[width=1\textwidth]{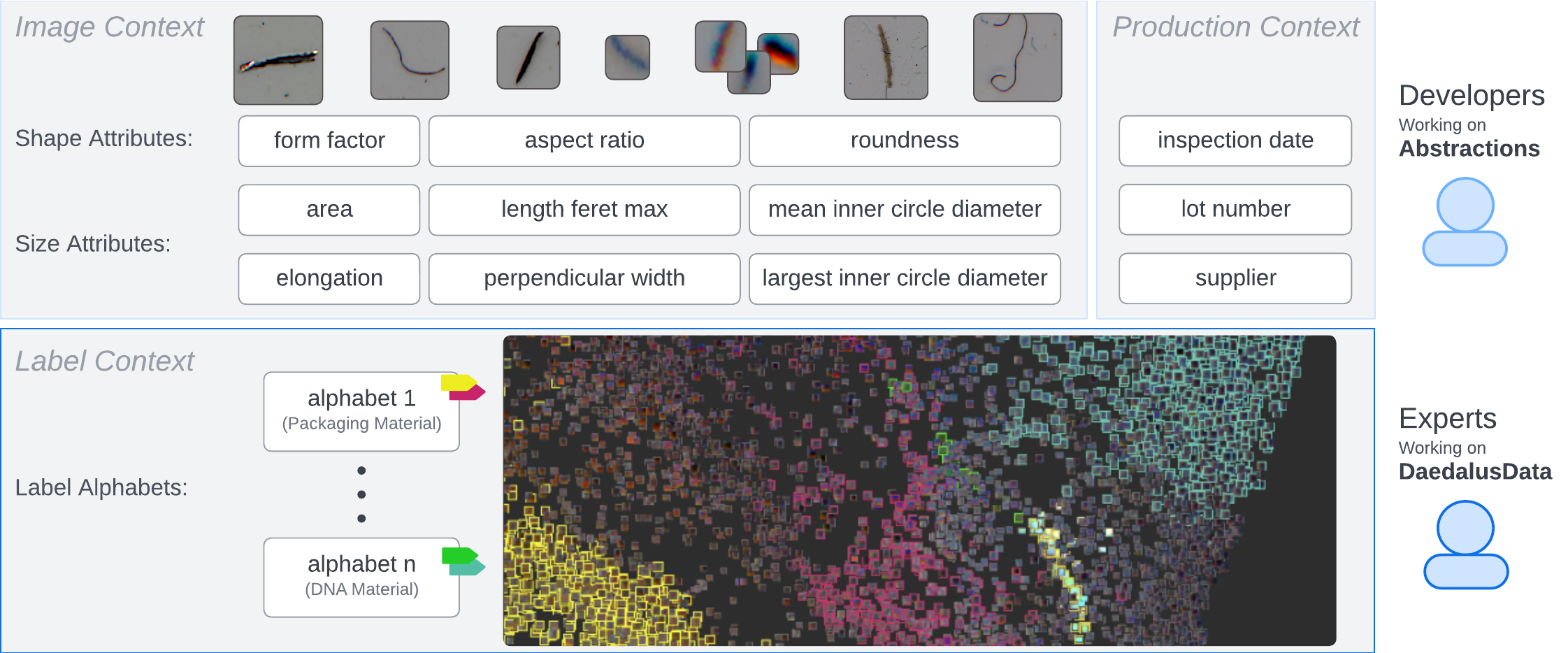}
    \caption{
    The overview of data abstractions reveals the added value of \toolname{}: while we carefully characterized 12 attributes during design (top), experts can further augment particle data through online interactive labeling (bottom). Multiple label alphabets can be re-played as augmented attributes, informing the positioning of particles for their semantic exploration.
    Top (design phase): The data abstraction included \emph{Image Context}, with nine numerical attributes on the \emph{size} and \emph{shape} of particles, computationally derived from images; in addition, the \emph{Production Context} offers three categorical/ordinal attributes contextualizing the particle origin.
    Bottom (application phase): the \emph{Label Context} is created by experts using \toolname{}, augmenting the dataset by labeling new attributes with rich domain semantics, expressed through multiple \emph{label alphabets}.
    }
    \label{fig:attributes}
\end{figure*}

\subsection{Process}
\label{sec:design_methodology}
Our methodology follows a design study~\cite{DBLP:journals/tvcg/SedlmairMM12} and is grounded in a tight collaboration between VA researchers of the University of Zürich, and Roche's product quality engineers responsible for a variety of consumables and particle contamination issues. 
Building upon the relationship between one of our researchers and this Roche team, communication, buy-in for the project, and access to the dataset were well-established from the start. 
We conducted a literature review and an exploratory interview~\cite{blandford2013semistructured} with a key domain expert to establish the existing gaps in detailed particle data analysis and to understand the current state of VA research in this space. 
Our interviewee from the learning phase (E1) is a co-author on this paper, as is common in design study methodology.

We consulted a diverse group of experts at Roche, including S1, a senior scientist from Roche Pharma Technical Development; E2-E4, three product quality engineers from Roche Diagnostics Global Operations Consumables;
a senior engineer experienced with different types of particle contamination from Roche Diagnostics Product Care (E5);
an R\&D scientist from Roche Pharma Particle \& Formulation Characterization (E6);
and two engineering team leads from Roche Diagnostics Global Operations Consumables \& Digital Excellence (M1-M2).
M2 served as gatekeeper~\cite{DBLP:journals/tvcg/SedlmairMM12}, being a member of the approval board for this project.
M1 and M2 acted as promoters~\cite{crisanFT21}, helped characterize the problem, serving as an invaluable development advisor within Roche, and steering the project's direction.
S1 acted as connector~\cite{DBLP:journals/tvcg/SedlmairMM12}, gatekeeper~\cite{DBLP:journals/tvcg/SedlmairMM12}.
E1-E3 served as front-line engineers~\cite{DBLP:journals/tvcg/SedlmairMM12}, each with unique expertise in assessing particle contamination in different consumables—vital for nuanced analysis requirements. 
E1, acting as translator~\cite{DBLP:journals/tvcg/SedlmairMM12} and our data steward~\cite{crisanFT21}, also contributed significantly to the problem characterization.

The project was completed over a 12-month period of iterative development, prototyping, expert feedback, refinement, and evaluation. 
We developed an initial set of requirements in Months 1-2 with E1 and M1, in parallel with the data abstraction process (result shown in Section~\ref{sec:data_characterization}).
E1 explained their workflow to us, resulting in Flow 1, data exploration to labeling, and its extension into Flow 2, for knowledge externalization, as described in Figure~\ref{fig:vis_concept} and Section~\ref{subsec:system_overview}.
After an initial development round in Months 2-4, we conducted a formative user study with E1, to evaluate our characterization of requirements, as well as the mental load of and user satisfaction with our prototype. 
The main design, implementation, and refinement phases that followed in Months 5-8 involved a highly iterative process.
We refer interested readers to the Supplementary Material, where we present intermediate outcomes of major iteration cycles in rich detail, and reflect on design choices. 
We conducted an additional user study with E1-E5, the results can be found in Section~\ref{sec:evaluation:userStudyResults}. 
We also present two case studies with E1 and M1, conducted in Months 11-12, to further validate \toolname{}.
Currently, \toolname{} is available within Roche as a proof-of-concept application.

\subsection{Domain Problem}
Quality control at Roche is a critical step in ensuring the safety and efficacy of consumables in diagnostic testing.
The current inspection process of consumables involves different tests, dependent on the criticality of the consumable, whereas visual inspections belong to the standard test procedure.
Abnormalities found during the inspection of a production lot, such as particle contaminants, are assessed based on type, size, and quantity regarding the product specification.
This task requires a trained eye and expertise since the particle evaluation can be ambiguous, hence the severity assessments can differ depending on the analysis context.
The main problems of experts are:

\begin{itemize}[noitemsep,leftmargin=*]
    \item The  amount of particles hinders experts from grasping visual patterns from the particle images, and performing labeling tasks efficiently.
    \item Particles contain multiple important attributes, but data-driven decision-making so far was only based on subsets of size and shape metrics, depending on the involved expert, task, and time budget.
    \item During expert inspection, valuable knowledge about particle characteristics is not systematically captured and utilized through labeling. Potential is lost for structuring particles by different categorizations (label alphabets), and for more collaborative assessments. 
\end{itemize}

A system that effectively supports exploration, effective labeling, and ''fluent'' knowledge externalization would enable more objective, data-driven decision-making around handling particle contaminants. 
Then, the domain experts can systematically categorize particles into meaningful classes, such as material types, streamlining their quality assessment process, leading to more consistent and efficient outcomes, and ultimately saving time and costs. 

\subsection{Data Abstraction}
\label{sec:data_characterization}

In our study, we gathered a dataset comprising 37,857 images of individual particles.
The particles were collected from 70 separate production lots of IVD consumables.
A production \emph{lot} refers to a batch of consumables produced during a specific period, under the same conditions, and intended to have uniform characteristics and quality.
The number of contaminating particles in a production lot influences whether the lot can be used for diagnostics, or should rather be disposed of entirely.
These consumables are manufactured by eight specialized suppliers, responsible for ensuring that their production lots meet the stringent quality standards required for IVD consumables. 
To maintain confidentiality and protect proprietary information, the data presented in this paper has been anonymized, including all dates, suppliers, and lot identifiers. 
Particles from the same lot are still grouped together.

Each particle image is processed through a proprietary object recognition algorithm, designed to identify, quantify, and measure particles within larger filter images, akin to public methodologies~\cite{RAADNUI2005871, DBLP:journals/tmi/ZhuCMP03}. 
In collaboration with domain experts, we distilled from 15 categorical/ordinal and 27 numerical attributes a set of 2 ordinal, 1 categorical, and 9 numerical attributes that provide a rich particle description, as depicted in Figure~\ref{fig:attributes}.
The 3 categorical/ordinal attributes represent the production context, while the numerical attributes, derived from the image, are subdivided into 3 shape and 6 size attributes.
All attributes are vital for identifying anomalies and tracing the origins of the consumables, informing the understanding of production conditions and potential contamination sources.
The size and shape attributes quantify the physical characteristics of particles, revealing a positive skew indicating a predominance of smaller particles.
The production attributes offer critical context for each particle, essential for comparing lots and suppliers.

Our images come preprocessed from our collaborators with 9 associated attributes, as described in Figure~\ref{fig:attributes}. 
We prepare these for \toolname{} by creating a feature vector, normalizing values for the numerical attributes, and one-hot encoding for the production context attributes.
This preprocessing ensures that the data is useful for analysis within the \toolname{} approach (details on our preprocessing methods are available in the Supplemental Materials).
\toolname{} utilizes the data mentioned above and augments the dataset with expert-developed label alphabets and labels, further explained in Section~\ref{labeling}.

\subsection{Requirements Abstraction}
\label{requirements}

Our iterative discussions with experts resulted in the identification of the requirements for \toolname{}, each reflecting an essential need experts have for enhancing the quality control process.

\paragraph{R1 — Attribute-Based Views:} It is essential for experts to partition the particle data based on attributes of interest.
The solution should facilitate this by clearly delineating particles, according to attribute categories and discrete bins for numerical attributes.

\paragraph{R2 — Comprehensive Overview:} \toolname{} must offer an overview of all particles, to reveal structural characteristics and provide experts with a clear understanding of what is present in the dataset.

\paragraph{R3 — Particle Filtering:} Experts need to drill down in the search space by attribute-based filtering. The tool must support the application of multiple, persistent filters for in-depth attribute-based exploration.

\paragraph{R4 - Visual Customization:} Experts need to adjust the visual representation of particle images for task-oriented display, having control over their relative and absolute size and background transparency.

\paragraph{R5 — Particle Selection and Inspection:} Experts need to select subsets of particles for detailed examination.
The solution should support this analysis by maintaining selections across different views and providing detailed information on a particle selection.

\paragraph{R6 — Label Alphabet Development:} Experts require externalizing their domain knowledge and insights gained during particle analysis.
The solution should allow experts to define multiple knowledge-informed label alphabets, depending on the dynamics of the task at hand; to be shared for collaborative use downstream.

\paragraph{R7 — Label-Informed Particle Display:} Experts need to structure particle images visually, according to particle characteristics, assigned labels, and both.
System adaptivity in this iterative process would enable dynamic particle visualizations tailored to specific tasks, for the discovery of task oriented patterns and relations in the data.

\paragraph{R8 — Efficient Labeling:} The solution needs to support persistent particle labeling for multiple label alphabets.
Labeling multiple selected particles at once will increase the efficiency of the process.

\begin{figure}[t]
  \centering
  \vspace{-0.5em}
  \includegraphics[width=\linewidth]{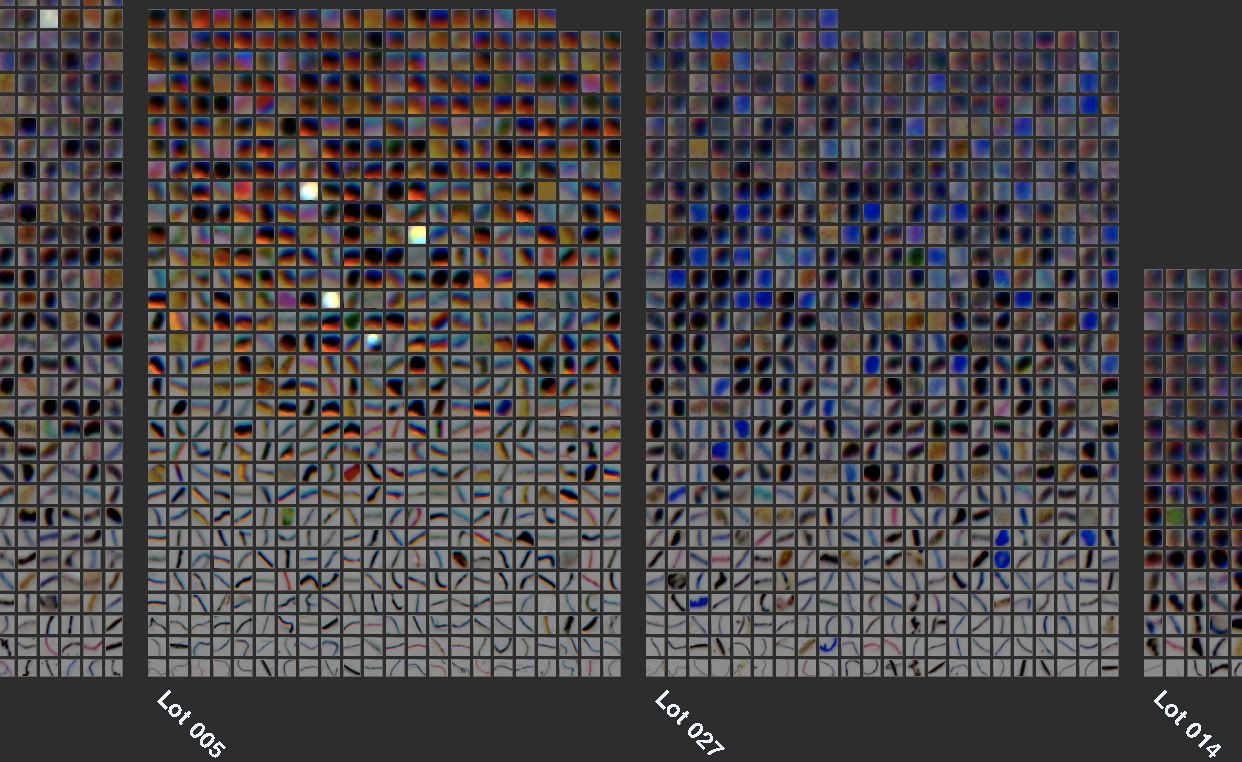}
  \caption{The Attribute View provides an overview of particles (R\textsubscript{1}), structured by a user-selectable attribute, easing the comparison of particles between attribute levels (categories, bins for numerical attributes). Here, the expert chose the \emph{Lot Number} and zoomed toward two lots, for a detailed inspection. The expert makes an interesting observation: they identified that many of the 669 particles of \emph{Lot 027} (right) appear to be blue, compared to the orange tone of \emph{Lot 005} (left).}
  \vspace{-0.5em}
  \label{fig:lot_column}
\end{figure}


\section{The \toolname{} Interface}
\label{sec:proposed_solution}
\toolname{} is a VA system designed to address complex analysis challenges of particle contamination of IVD consumables.
We provide a system overview, and our design rationale for data exploration, knowledge externalization, and labeling. 

\subsection{System Overview}
\label{subsec:system_overview}
\toolname{}'s design is centered around the display of thousands of particle images.
Experts call this view the \emph{Canvas}, the starting point for detecting areas of interest.
We contribute a conceptual underpinning that eases generating different types of particle displays, informed by different requirements, as Figure~\ref{fig:vis_concept} shows.
Two types of control enable experts to influence the positioning of particles on the Canvas:
\begin{itemize}[noitemsep,leftmargin=*]
    \item \textbf{Single vs. Multi-attribute:} structuring particles by a single data attribute, or projecting particles by multiple attributes. This enables experts to explore particles from different perspectives, reveal areas of interest, and eases the labeling (Figure~\ref{fig:vis_concept} top vs. bottom)
    \item \textbf{Pre-existing vs. Augmented attributes:} structuring particles using the initially abstracted data attributes, and/or using expert-provided label alphabets as attributes, extending the exploration and labeling workflow through knowledge externalization (Figure~\ref{fig:vis_concept} left vs. right)
\end{itemize}

As Figure~\ref{fig:vis_concept} suggests, \toolname{} implements the full cross-cut of these two types of control, offering high flexibility for experts in structuring particles in the Canvas.
The \textit{Attribute View} enables partitioning of particles by a \textit{single attribute}, crucial for inspecting and comparing various groups within the dataset (R\textsubscript{1}). 
The \textit{Projection View} enables dimensionality-reduced projections of particles, based on any combination of user-specified \textit{multi-attribute} set.
This similarity-preserving particle display enables the identification of significant patterns or anomalies within the particle data (R\textsubscript{2}).
Both views accept \textit{pre-existing attributes}, but also enable users to explore \textit{augmented attributes} (Figure~\ref{fig:attributes} top vs. bottom).
Our data augmentation approach treats expert-specified labels and label alphabets as attributes, enabling experts to generate insights using their externalized knowledge (R\textsubscript{7}).

\toolname{}'s Canvas also includes support for robust data exploration through its Auxiliary Views (Figure~\ref{fig:vis_concept}, right). 
The \textit{Filter View} (Figure~\ref{fig:filter}) includes attribute-based filters, so experts can refine the dataset and observe the effects of applying multiple filters on the Canvas (R\textsubscript{3}).
The \textit{Selection-Inspection View} (Figure~\ref{fig:view_f}) reveals summary statistics on particles currently selected by the expert on the Canvas (R\textsubscript{5}). 
The Detail View (Figure~\ref{fig:filter}) displays an enlarged image of a single expert-selected particle, along with all of its image and production context attributes (R\textsubscript{5}).
Finally, the \textit{Label View} (Figure~\ref{fig:add_label}) allows experts to define, select, and modify label alphabets (R\textsubscript{6}), translating domain knowledge into actionable labels (R\textsubscript{8}). 

\toolname{} supports two workflows, informed by iterative work with the experts, reflected in Case Studies 1 and 2 (Section \ref{sec:evaluation:caseStudies}).

\begin{enumerate}
    \item \textbf{Data Exploration for Labeling:} To label a group of particles, experts engage in an iterative cycle of exploring the dataset by individual (R\textsubscript{1}) or all data attributes to reveal structural characteristics (R\textsubscript{2}), applying attribute-based filters (R\textsubscript{3}), and customizing their visual representation (R\textsubscript{4}). After generating insights on a selected particle group (R\textsubscript{5}), experts assign a label to it (R\textsubscript{8}).
    \item \textbf{Knowledge Externalization for Labeling:} To label a group of particles, experts extend the data exploration cycle by continually contributing labels to the dataset, which they then further organize into label alphabets (R\textsubscript{6}). They then compute projections with these label alphabets, resulting in informed projections (R\textsubscript{7}), leading to the discovery of new relationships within the data and further externalization of domain knowledge.
\end{enumerate}

\begin{figure}[t]
  \centering
  \vspace{-0.5em}
  \includegraphics[width=\linewidth]{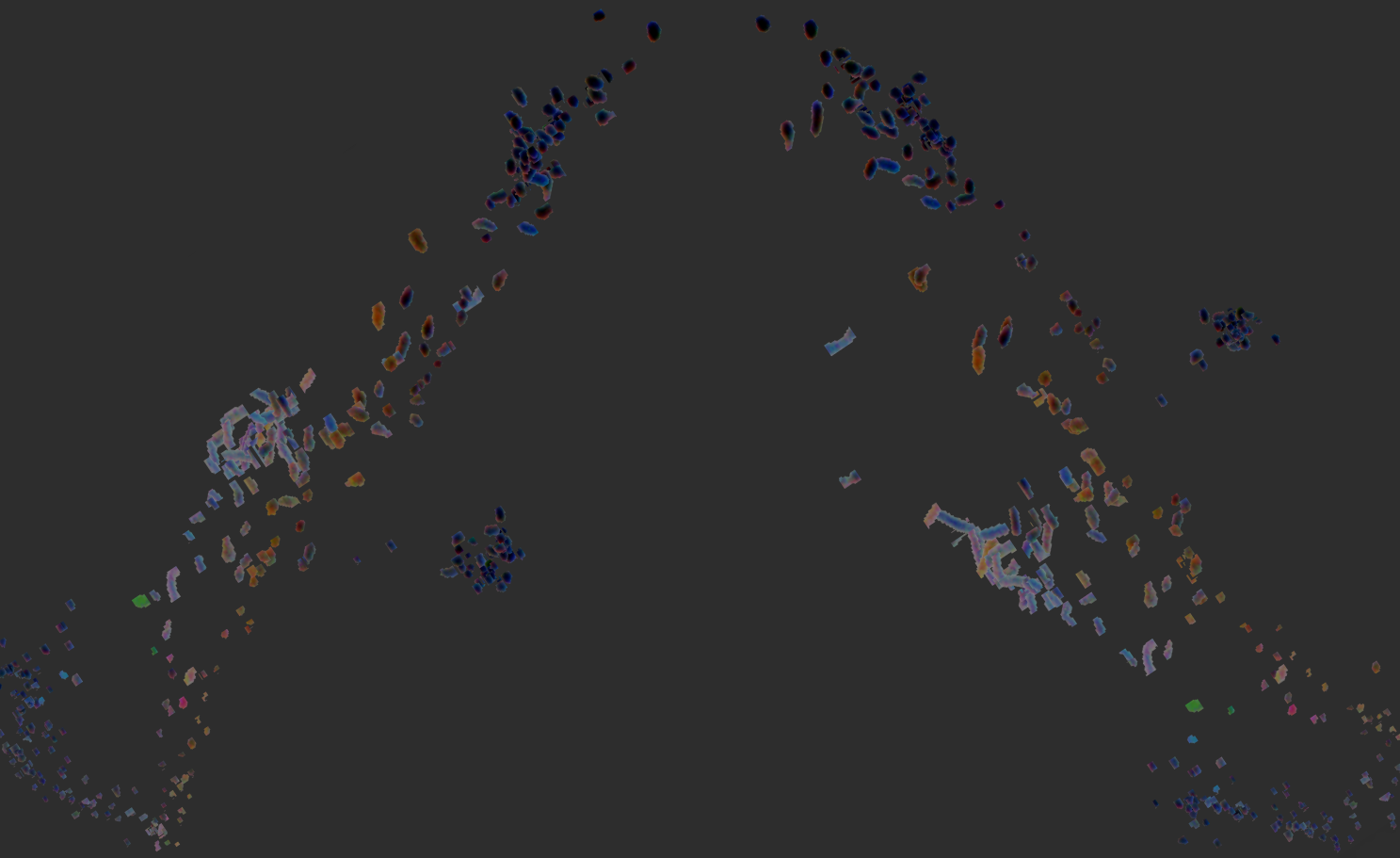}
  \caption{%
  The \emph{Projection View} provides a comprehensive similarity-preserving particle overview (R\textsubscript{2}), leveraging dimensionality reduction.
  Experts have full flexibility in the selection of attributes, triggering re-computation.
  Adding an augmented expert-labeled attribute creates a \emph{Label-Informed Projection}, representing added domain semantics (R\textsubscript{7}).
  In this example, the expert combined image attributes with the \emph{Lot Number}.
  The projection shows an almost symmetric structure (left vs. right), representing two lots of high similarity opposite each other.
  }
  \vspace{-0.5em}
  \label{fig:lot_similarity}
\end{figure}

\subsection{Particle Canvas and Canvas Interaction}
The \textit{Canvas} depicts particle images as small thumbnails positioned in x and y by leveraging either the Attribute View or the Projection View technique (R\textsubscript{1}, R\textsubscript{2}).
To increase the visual scalability to many thousands of particles, experts can zoom in for a detailed investigation of specific particles and use panning to navigate across the Canvas.
Auxiliary Views support filtering and other multi-particle interactions (R\textsubscript{2}, Sections~\ref{filtering}-~\ref{labeling}), with effects propagated back to the Canvas.
Experts can interact with controls on the left, to steer the vertical dimension of our framework:
Single-attribute selection opens the Attribute View (Figure~\ref{fig:lot_column}), and multi-attribute selections open the Projections View (R\textsubscript{2}) (Figure~\ref{fig:lot_similarity}).

\begin{figure}[t]
  \centering
  \vspace{-0.5em}
  \begin{subfigure}{.32\linewidth}
    \centering
    \includegraphics[width=\linewidth]{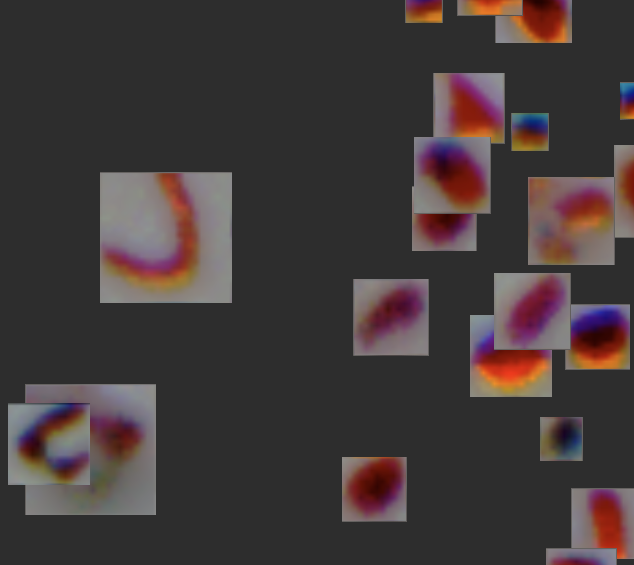}
    \caption{Non-uniform size and non-transparent particle images}
    \label{fig:e_uniform_non_transparent}
  \end{subfigure}%
  \hfill
  \begin{subfigure}{.32\linewidth}
    \centering
    \includegraphics[width=\linewidth]{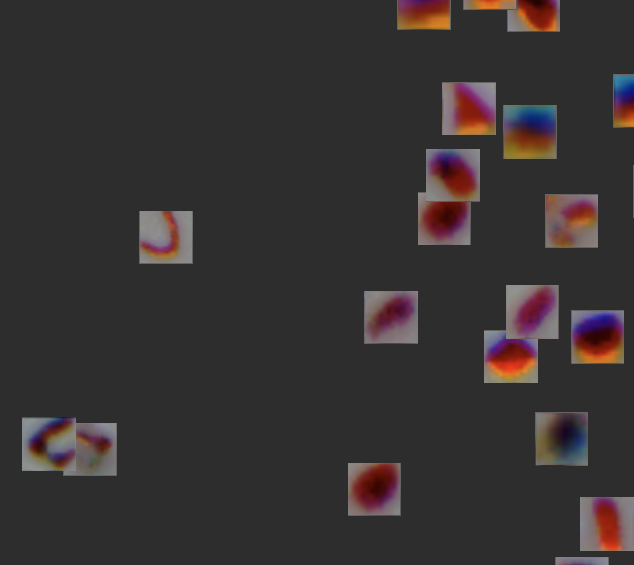}
    \caption{Uniform size but non-transparent particle images}
    \label{fig:e_uniform}
  \end{subfigure}%
  \hfill
  \begin{subfigure}{.32\linewidth}
    \centering
    \includegraphics[width=\linewidth]{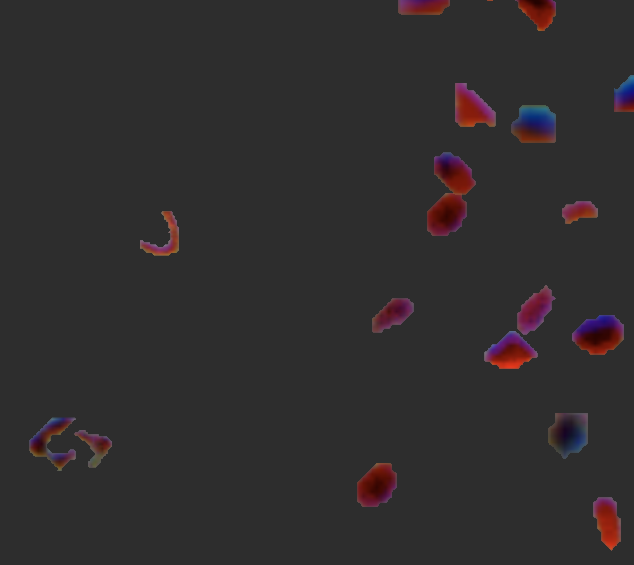}
    \caption{Uniform size and transparent particle images}
    \label{fig:e_uniform_transparent}
  \end{subfigure}
  \caption{Visual customization in practice (R\textsubscript{4}). 
  These three examples show possible Canvas customization support within \toolname{}(R\textsubscript{4}).
  Customization helps experts mitigate overplotting: experts can declutter their Canvas by toggling between relative and absolute particle sizes, and calibrating how much the image background is obscured.
  }
  \vspace{-0.5em}
  \label{fig:view_e}
\end{figure}

\toolname{}'s Canvas helps experts address visualization issues like overplotting (R\textsubscript{4}).
Experts can configure the positioning, size, and transparency of particle images, tailoring the data presentation to their analysis needs to enhance visibility of patterns and relationships among particles.
Figure~\ref{fig:view_e} demonstrates the effects of:
\begin{itemize}[noitemsep,leftmargin=*]
    \item \textbf{Uniform Size:} Standardizes particle image size for relative comparisons (Figure~\ref{fig:e_uniform}).
    \item \textbf{Transparent Objects:}Renders image backgrounds transparent, focusing on particles (Figure~\ref{fig:e_uniform_transparent}).
\end{itemize}

\subsection{Attribute-Based Particle Exploration}
The Attribute View enables experts to explore large numbers of particles by a selected attribute of interest (R\textsubscript{1}, Figure~\ref{fig:lot_column}).
Aligning with the horizontal dimension of our framework, it accepts both it accepts pre-existing and label-based attributes.
Particle images are positioned in columns within nested grids, enabling experts to quickly compare and analyze particle distributions.
For numerical attributes, we use domain-preserving binning~\cite{muhlbacher2013partition} to arrive at discrete columns.
Together with the experts, we decided that the \emph{Elongation} attribute always dictates the vertical stacking order of particles, leading to a natural and semantically rich particle ordering, with the largest particles at the bottom.
In Figure~\ref{fig:lot_column}, an expert has, e.g., selected the \emph{Lot Number} attribute, enabling the effective comparison of particles of two lots in high detail.
The Attribute View can also highlight particles, e.g., with specific labels, helping experts inspect label distributions across attribute categories.

\subsection{Projection-Based Particle Exploration}
\label{projection_exploration}
The Projection View provides an overview of thousands of particles (R\textsubscript{2}) by projecting similar particles close to each other.
This aids exploratory analysis and subsequent decision-making in quality control, by helping to identify particle patterns or anomalies (see Figure~\ref{fig:lot_similarity}).
We use semi-supervised UMAP~\cite{mcinnes2018umap-software} for dimensionality reduction, on both pre-existing (R\textsubscript{2}) (image and production context) and label-based attribute projections, according to the horizontal framework dimension.
Experts can select attributes and compute new projections on the fly, allowing dynamically changing perspectives of interest.

For label-based attributes, the Projection View acts as a \textit{Label-Informed Projection} (Figure~\ref{fig:view_f}), incorporating semantic information from labeled data (R\textsubscript{7}), supporting the Knowledge Externalization Workflow.
This creates custom projections for specific analytical questions, e.g., about particle relationships based on label alphabets of interest.
Label-informed projections can be updated as more particles are labeled, providing a dynamic view that evolves with the analysis.

\subsection{Particle Filtering}
\label{filtering}
\begin{figure}[t]
  \centering
  \vspace{-0.5em}
  \includegraphics[width=\linewidth]{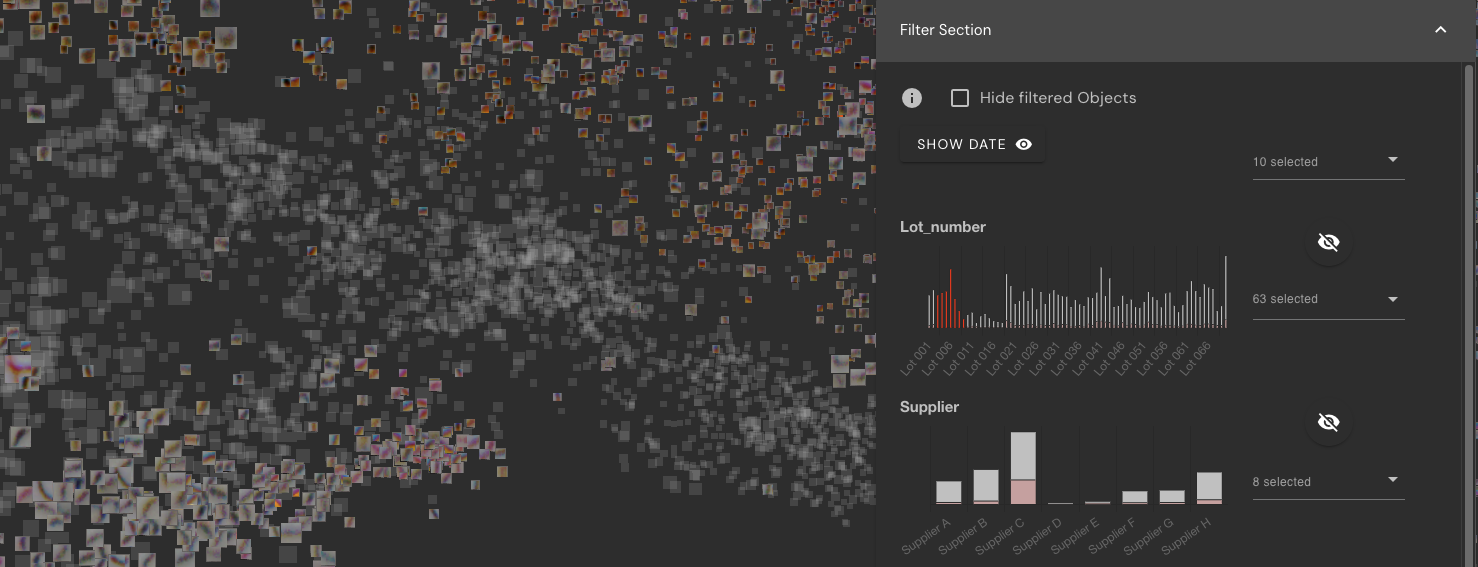}
  \caption{
  \emph{Projection View} of thousands of images, with the \emph{Filter View} (R\textsubscript{3}) on the right.
  Expert-selected filters for \emph{Lot Number} (top) and \emph{Supplier} (bottom) narrow down the search space.
  The Canvas encodes which particles have been filtered out in the projection by coloring their images light gray. 
  In dark gray, the bar chart encodes the exclusion of suppliers ''D'', ''E'', ''F'', ''G''.  
  In red, the bars encode the number of particles filtered by the other attribute, alerting the expert to possible relationships between selected filters. 
  }
  \vspace{-0.5em}
  \label{fig:filter}
\end{figure}

Experts can use the Filter View (Figure~\ref{fig:filter}, right) to narrow down the particle dataset for more focused analysis (R\textsubscript{3}).
The filter interactions consider pre-existing \textit{and} label attributes as equally valid, consistent with \toolname{}' horizontal framework dimension and data augmentation approach.
Experts can access attribute-based statistical summaries and visualizations of the filtered dataset, supporting exploration, insight generation, and efficient labeling.

The Filter View, aside from filtering controls, displays the particle data distribution of each filter attribute as a stacked bar chart.
To preserve context, visualizations show the entire particle dataset, with particles included and excluded by the filter encoded as light gray and red bars, respectively.
When multiple filters are selected, each bar encodes the number of particles filtered by other attributes in red, highlighting possible filter relationships.
Filtering effects directly impact the Canvas, providing cleaner, more focused displays.
Excluded particles are shown as gray background boxes on the Canvas (Figure~\ref{fig:filter}) and can be hidden entirely from the Canvas.

\subsection{Detailed Particle Inspection}
\label{inspection}
\begin{figure}[t]
  \centering
  \includegraphics[width=\linewidth]{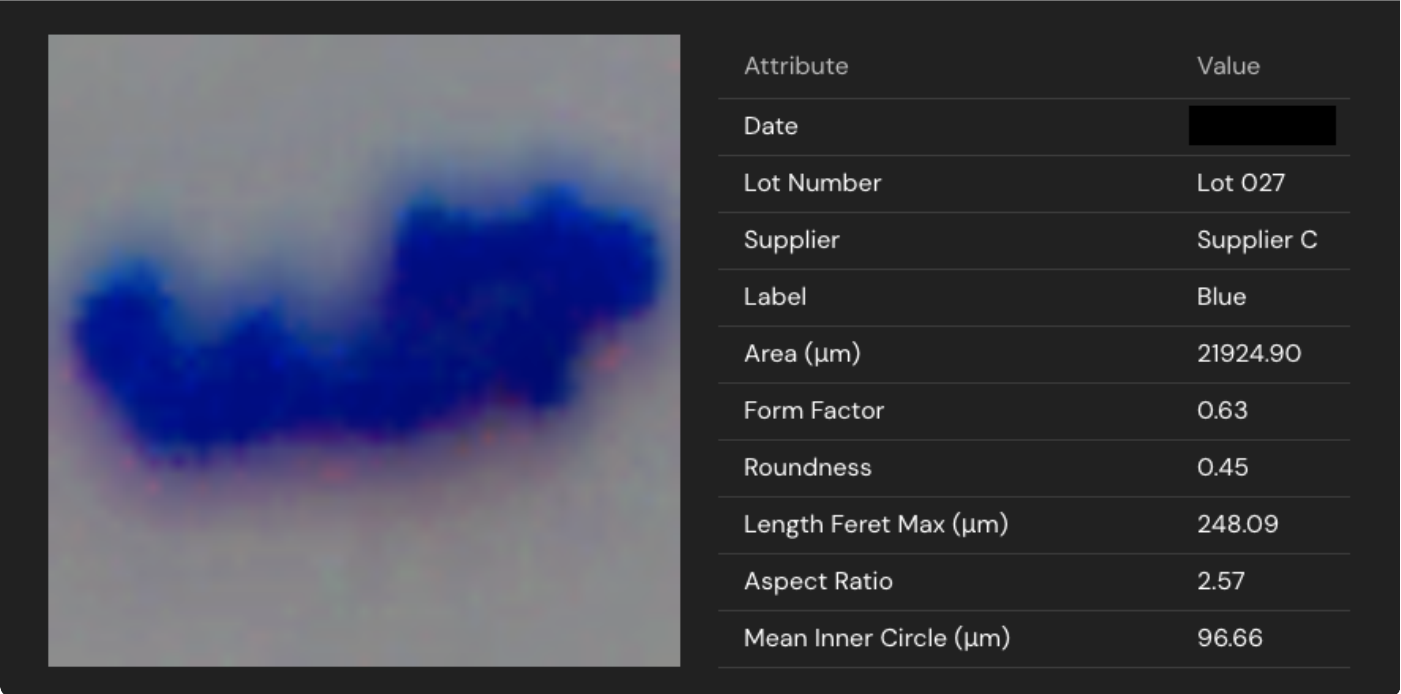}
  \caption{\emph{Detail view} of a single selected particle (R\textsubscript{5}). 
  Experts can analyze the particle image in detail and retrieve specific information for all 12 abstracted attributes, informing data labeling and further analyses downstream.
  Here, a blue particle was examined, referring to \emph{Lot Number} 27; this was the result of a detail-on-demand selection in the \emph{Attribute View}, as shown in Figure~\ref{fig:lot_column}. 
  } 
  \label{fig:detail_view}
  \vspace{-0.5em}
\end{figure}

\begin{figure}[t]
  \centering
  \includegraphics[width=\linewidth]{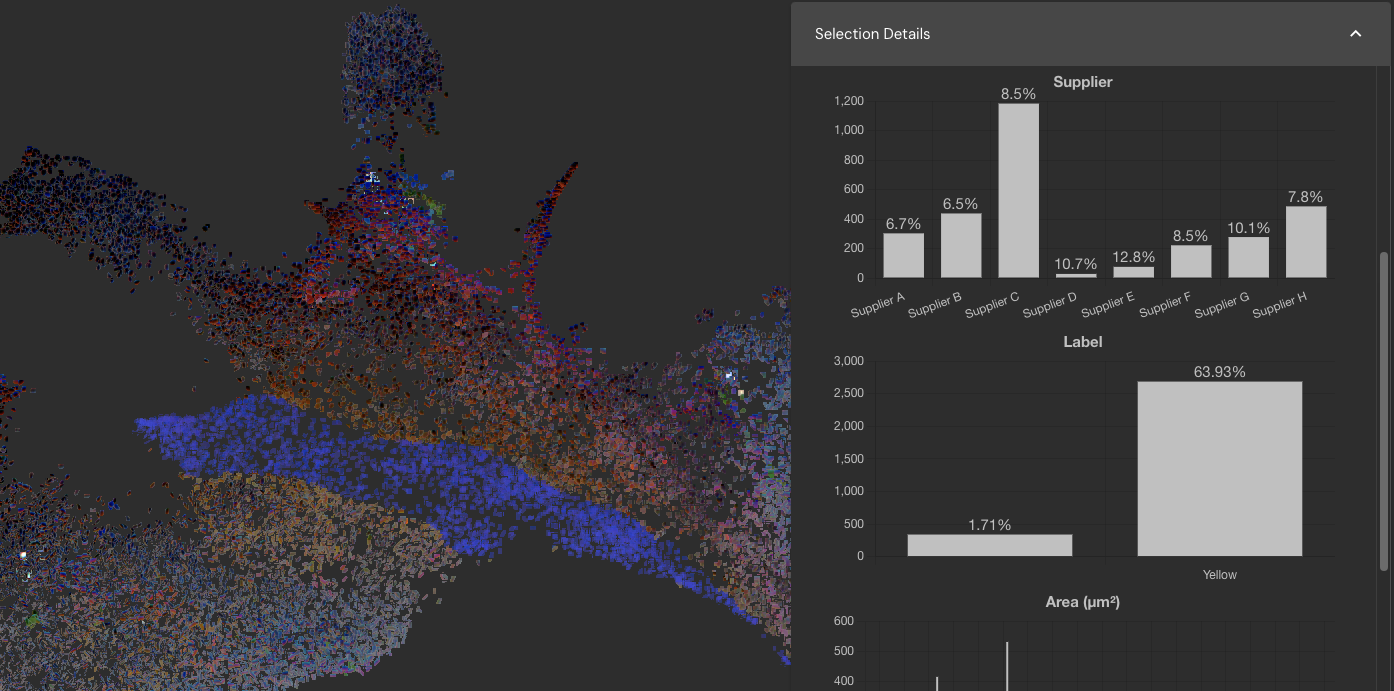}
  \caption{
  The \emph{Selection-Inspection View} on the right helps to contextualize the multi-particle selection (R\textsubscript{5}) made in the \emph{Label-Informed Projection} (R\textsubscript{7}) on the left.
  The per-attribute percentage values above each bar account for relative particle frequency.
  Here, selected particles are represented the most by Supplier C (10.2\%), and are overwhelmingly labeled by experts as ''Yellow''(74.26\%). 
  Hovering on one of these bars reveals a tooltip showing the number of unlabeled particles.
  }
  \vspace{-1.0em}
  \label{fig:view_f}
\end{figure}

\begin{figure}[t]
  \centering
  \includegraphics[width=\linewidth]{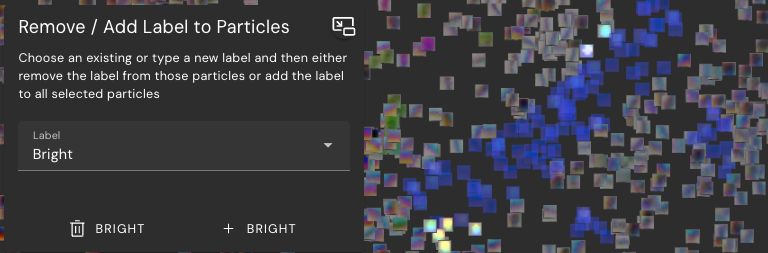}
  \caption{Projection View (right) with a group of expert-selected particles appearing in blue on the Canvas.
  The Label View (left) shows that the expert has selected the label ''Bright'', and can now either add or remove this label from their selection (bottom left and right).}
  \label{fig:add_label}
  \vspace{-2mm}
\end{figure}

For rapid access to information on selected particles, experts can refer to the \textit{Detail View} and the \textit{Selection-Inspection View}.
These views provide details on the attribute values for selected particles (R\textsubscript{5}), allowing for thorough examination of particles of interest.
This informs experts on anomalies or patterns, generating insights for decision-making and creating new labels. 
Clicking on a single particle or a selected group in the Canvas triggers these on-demand views.
The Detail View shows all attributes for a single particle, along with an enlarged particle image (Figure~\ref{fig:detail_view}).
The Selection-Inspection View offers attribute-specific charts displaying the distribution of the selected particle group within the dataset (Figure~\ref{fig:view_f}).
The lasso and rectangle tools enable bulk selection, selected particles are visually encoded with a blue "glow": a blue-transparent area(Figure~\ref{fig:add_label}).
In this way, particles added to a selection are visually modified.
All selected particles are summarized by their attribute value distributions in the Selection-Inspection View (Figure~\ref{fig:view_f}).

\subsection{Label View}
\label{labeling}

\begin{figure}[t]
  \centering
  \includegraphics[width=\linewidth]{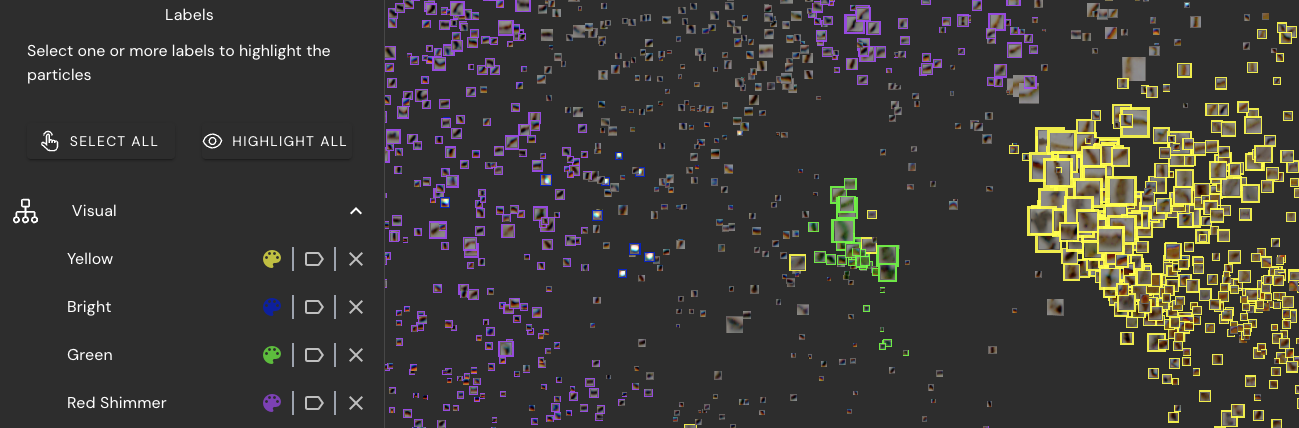}
  \caption {Projection View (centered), with particles highlighted by labels.
  The Label View (left) shows the 'Size' alphabet, with expert-assigned colors per label.
  These colors help to distinguish pixels in the Canvas, highlighted by their label color.}
  \label{fig:label_highlight}
  \vspace{-2mm}
\end{figure}

Experts use the \textit{Label View} (Figure~\ref{fig:label_highlight}) to determine which particles are associated with each label and to create and manage label alphabets.
Label alphabets are groups of at least one label, representing the expert's perspective on the relationship between several labels (R\textsubscript{6}).
Label alphabets provide structure to the labels specified by the expert and enhance the knowledge externalization workflow in \toolname{}.
They also enable experts to create label-informed projections (Section~\ref{projection_exploration}), offering insights into particle contamination based on labels, alphabets, and their relationships, otherwise impossible for experts to express. 

For the effective development and extension of label alphabets, precise multi-selection tools for similar particles are key (R\textsubscript{8}). 
The Label View's visual encoding includes symbols for controlling the Canvas and adding descriptions to existing labels.
Experts can select multiple labels, visually encoding particles with a blue glow, and triggering particle data summaries in the Selection-Inspection View.
Highlighting multiple labels encodes particle images with outlines of distinct colors per label, maintaining both label indication and particle visibility (R\textsubscript{5}).

\subsection{Implementation}

\toolname{} is a single-page application written in Typescript using the Vue3 framework.
The visualizations, including particle representation on the Canvas, are built using Three.js with WebGL shaders~\cite{danchilla2012three, angel2014interactive} and ChartJS~\cite{chartjs}.
Communication between the frontend and backend is implemented using a GraphQL interface.
The backend is a Python FastAPI application that manages the dimensionality reduction calculation (UMAP)~\cite{mcinnes2018umap-software} and validates database interactions (Figure S16 in the Supplemental Materials).
A MongoDB NoSQL database stores the particle labels and their images, size, shape, and inspection attributes for recalculation.
A Postgres relational database stores the labels and label alphabets, and all particle coordinates within a projection.


\section{Evaluation}
\label{sec:evaluation}

We evaluated the usefulness of \toolname{} in two case studies with front-line engineers, with familiarity with a particle dataset at hand.
In addition, we conducted a user study to assess the usability, with an experiment involving \toolname{}'s requirements.

\subsection{Case Studies}
\label{sec:evaluation:caseStudies}
We present two case studies, drawn from expert usage sessions (E1, M1), to illustrate the usefulness of \toolname{}.
We use \videotimestamp{V, 03:36} as a code, referring to the supplemental video, for the reader's convenience.

\subsubsection{Case Study 1: Data Exploration and Labeling}
\label{sec:evaluation:caseStudies1}

This case study was performed by E1, a front-line engineer interested in gaining insights on an anomaly reported in Lot 27, characterized by an unusually high number of blue particles, which have previously been associated with changes in work clothing and cleaning rags that linted.
To gain an overview of the particle data, E1 used the Attribute View, to partition by lot number (\videotimestamp{V, 03:41}).
E1 could easily confirm the appearance of this abnormal number of blue particles for Lot 27.

Informed by this anomaly, E1 wondered if other lots might also be affected by blue particles. 
Given the large number of 70 lots, all with different ratios of blue particles, E1 switches to the Projection View for a more holistic and similarity-preserving view.
Guided by the already highlighted blue particles, E1 identifies a dense area with more unlabeled, blue particles from other lots.
E1 identifies the dense area and applies zooming for enlarged particle display (\videotimestamp{V, 04:13}).
A lasso selection leads to the efficient selection of hundreds of blue particles with only a single interaction.
To avoid false positives, E1 zooms closer, and deselects particles of other color, on a per-case basis, leading to a final selection of particles with 100\% blue color.
Happy with the selection, E1 is ready to set a label (''blue''), which will persist this particle selection, also for later use.

Coming back to his concern about other lots with blue particles, E1 uses the Selection Detail View for the contextualization of selected particles (\videotimestamp{V, 04:52}).
The result of the analysis is clear: in none of the remaining lots is evidence for an extraordinary degree of blue particles; a very positive outcome: all remaining lots run under normal condition.

In the process of labeling the blue particles, E1 spotted some ''bright'' or ''reflective'' particles in the Attribute View (\videotimestamp{V, 04:35}).
After completing the labeling of blue particles, E1 identified a larger group of bright particles in the resulting Projection View.
E1 found it helpful to switch between more general and more specialized projections during this exploration process, as this revealed even more new patterns and potential groupings of particles they could label, beyond E1's expectations.

\subsubsection{Case Study 2: Knowledge Externalization and Labeling}
\label{sec:evaluation:caseStudies2}

M1 was interested in label validation and alphabet development, to increase labeling efficiency and to make future decisions more structured, especially regarding supplier management.
He started his exploration in \toolname{} with 40\% labels.
Notably: these labels have been created not by M1 and were previously unknown to him.
In the Label-Informed Projection using these labels as an attribute, he identifies an interesting star-like shape that gains his interest (\videotimestamp{V, 05:11}).
M1 identifies a dense region at the center and many surrounding spots with interesting structures.
Using the lasso for one of the spots, M1 quickly selected around 350 particles of interest.

Aiming to make sense of the selection, M1 switched to the Selection Details View, searching for attributes that explain the particle selection.
He identifies that almost all particles refer to the same \emph{Production Lot} 28.
Informed by this finding, he repeats the process for a second spot and confirms: that spots in the projection can be explained by the particles' \emph{Production Lot}.
M1 now turns to a more fine-grained analysis of these two spots/lots, using zooming and panning.
Figure~\ref{fig:lot_similarity} shows the result: two local particle structures of high symmetry become apparent.
M1 selects the particles of both spots, aiming for their detailed analysis.
Interestingly, the Selection Detail View reveals that all particles can be associated with the same supplier (\videotimestamp{V05:41}).
Coming back to the high similarity between lots observed earlier, M1 concludes that this supplier has made deliveries of high homogeneity, a very pleasant signal.

M1 continues with focusing on the 60\% of unlabeled data, and starts his exploration in the Projection View.
The patterns he identifies, including a combination of size and distinctly colored contaminants, make him confident that with \toolname{}, alphabet creation and labeling of many thousands of particles would be possible with an efficiency so far unseen.
Summarizing his utility experience, M1 suggests the collaborative use of \toolname{}, involving a highly skilled expert for alphabet creation, and one employee responsible for labeling operations.

\paragraph{Task-Based Assessment of Effectiveness}

\subsection{User Study}
\label{sec:evaluation:userStudy}
\subsubsection{Experiment Design}
\label{sec:evaluation:caseStudiesDesign}

We conducted a user study to capture a comprehensive understanding of \toolname{}'s usability and effectiveness from the perspective of domain experts.
The methodology emphasizes qualitative analysis, supplemented by task-specific benchmarks and a NASA Task Load Index, to gain insight into user interactions with \toolname{}.

We selected participants with high expertise, as is typical in design studies.
Participants were chosen based on their proficiency in quality control and their familiarity with particle contamination issues in IVD consumables, ensuring that the collected data and feedback were relevant and informed by practical experience in the field.
In the study, we involved five experts working at Roche Diagnostics in the quality domain, including four Product Quality Engineers and a Senior Engineer from system integration.
We conducted the study in a controlled setting at Roche in Rotkreuz (CH), where  participants used identical hardware setups to ensure a uniform user experience.
The study infrastructure operated locally on a provided laptop, to minimize external factors that could potentially affect study outcomes. 

\begin{table}[b]
    \centering
    \begin{tabular}{lcccc}
        \toprule
        Expert & Task 1 & Task 2 & Task 3 \\
        \midrule
        E1 & \cmark\cmark\cmark & \cmark\cmark & \cmark\cmark \\
        E2 & \cmark\cmark\xmark & \cmark\cmark & \cmark\cmark \\
        E3 & \cmark\cmark\cmark & \cmark\cmark & \cmark\cmark \\
        E4 & \cmark\cmark\cmark & \cmark\cmark & \cmark\cmark \\
        E5 & \cmark\cmark\cmark & \cmark\cmark & \cmark\cmark \\
        \midrule
        Completion & 89\% & 100\% & 100\% \\
        \bottomrule
    \end{tabular}
    \caption{Sub-task completion for Tasks 1-3 by experts E2-4. A filled circle indicates task success (criteria defined in the supplemental material). One sub-task of Task 1
    caused some difficulties for one participant, in finding multiple occurrences of green particles in a projection.
    }
    \label{tab:subtask_completion}
\end{table}

We structured the study sequentially, with the following phases: an introductory overview of \toolname{}, a task-based assessment, a think-aloud protocol, a post-task evaluation using the NASA Task Load Index, and a concluding debriefing session.
This approach allowed for a thorough evaluation of the tool's capabilities.
A detailed user study agenda can be found in the supplemental material.
The Behavioral Observation Research Interactive Software (BORIS) was used to categorize cognitive, physical, and verbal user behaviors~~\cite{https://doi.org/10.1111/2041-210X.12584}, to uncover user interaction patterns.
We analyzed this data and created event sequences of participant interactions, to study the usage of the Canvas and Auxiliary Views.
The task-based evaluation included nested sub-tasks to align with the requirements and the tool's intended use.
The main tasks are defined as follows, with detailed descriptions and acceptance criteria for sub-tasks provided in Table~\ref{tab:subtask_completion} and listed in Appendix S2.2:

\begin{itemize}[noitemsep,leftmargin=*]
    \item \textbf{Task 1: Data Exploration}, including particle identification and particle group identification
    \item \textbf{Task 2: Data Labeling}, including adding a label, labeling particles, and label review
    \item \textbf{Task 3: Knowledge Externalization}, including label-informed projections and Canvas navigation
\end{itemize}

Ethical considerations were taken into account throughout the study.
We obtained informed consent from all participants and maintained their confidentiality and anonymity.
We informed all participants about the data collection process and the intended use of the data, ensuring transparency and ethical integrity.

\subsubsection{Results}
\label{sec:evaluation:userStudyResults}

Table~\ref{tab:subtask_completion} shows the results of the task-based assessment.
Overall, the five experts succeeded in 97\% of the sub-tasks.
Our evaluation observed participants' "organic" completion of their analysis tasks, with only minimum interruption, allowing E2 to proceed after identifying a significant particle group.
The results underline \toolname{}'s usability and efficiency in identifying, patterns and extracting expert knowledge through labels.

\begin{table}[b]
    \centering
    \begin{tabular}{lccccccc}
        \toprule
        Expert & MD & PD & TD & Perf. & Eff. & Frustr. & Mean \\
        \midrule
        E1 & 50 & 5 & 10 & 20 & 20 & 25 & 22 \\
        E2 & 25 & 20 & 5 & 35 & 15 & 25 & 21 \\
        E3 & 30 & 25 & 50 & 65 & 20 & 25 & 36 \\
        E4 & 40 & 15 & 35 & 25 & 15 & 10 & 23 \\
        E5 & 35 & 15 & 15 & 30 & 40 & 50 & 31 \\
        \midrule
        Mean & 36 & 16 & 23 & 35 & 22 & 27 & 27 \\
        \bottomrule
    \end{tabular}
    \caption{NASA Task Load Index results from the user study. Abbreviations: MD (Mental Demand), PD (Physical Demand), TD (Temporal Demand), Perf. (Performance), Eff. (Effort), Frustr. (Frustration). All results are rounded to the nearest integer.}
    \label{tab:nasa_tlx}
\end{table}

\begin{figure}[b]
  \centering
  \includegraphics[width=\linewidth]{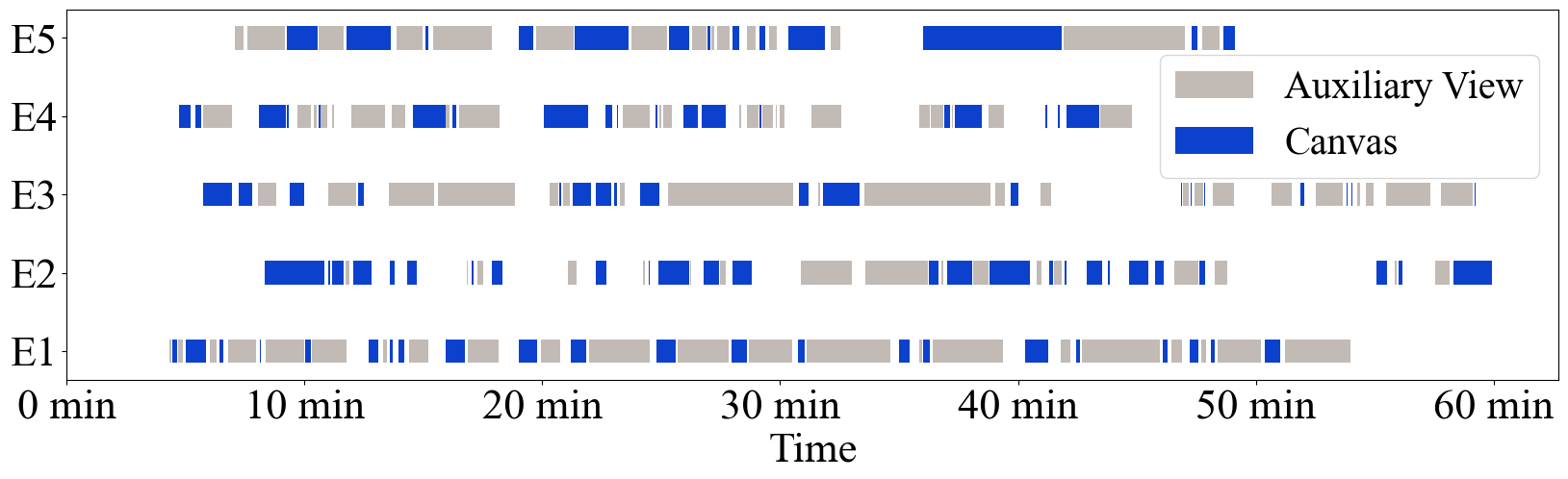}
  \caption {Annotated sequences showing the experts' interactions with \toolname{}, differentiating between Canvas (blue) and Auxiliary Views (gray) usage. Despite varying frequencies, all experts consistently utilized both Canvas and Auxiliary Views before switching back.
  }
  \label{fig:sequences}
\end{figure}

Figure~\ref{fig:sequences} provides insights into the time participants E1-E5 spent either with Canvas or Auxiliary View interaction during the task-based evaluation.
The five event sequences illustrate the diverse usage forms of \toolname{}, with E2 being rather Canvas-oriented, and E5 mainly focusing on Auxiliary Views.
What can also be seen is that the frequency of switching between the Auxiliary Views and the Canvas differs among experts.
It would be an interesting aspect for future work to study usage forms of \toolname{} more empirically.

\paragraph{Think-aloud Protocol} 

Participants found the \textbf{data exploration} enabled by \toolname{} to be highly usable and efficient, especially its intuitiveness for particle comparison.
E3 noted, ''The Attribute View is nice, with it, I can quickly compare between suppliers/lots''.
The tool's ability to maintain filters and selections when switching views was highlighted as a standout feature, enhancing workflow continuity.
E5 was positively surprised: ''I did not expect the filters and selection to persist when I switch between projections''.
Experts also expressed a need for more filtering options, including attributes like color and categorical shape (e.g., fibers).
While it is already possible to do this by labeling particles with these attributes, experts expressed interest in further integrating particle validation processes for confirming particle composition. 
E5 suggested expanding \toolname{} to include monitoring and trend analysis capabilities, leveraging physical inspection attributes.
E2 suggested integrating with external databases, such as internal resource planning systems, to enrich particle behavior analysis over time.

All participants appreciated the label-informed projection technique as an added value to their \textbf{knowledge externalization} process.
E5 noted: ''I really like the different algorithms and layouts (Attribute and Projection Views)'', providing different perspectives on particles.
The knowledge externalization support was considered an asset, potentially beneficial in other Roche-internal systems when combined with other particle-related data sources.

For particle \textbf{labeling}, two participants initially struggled with the lasso selection tool, preferring the rectangle selection tool instead.
However, over the course of the study, the benefit of having different means for particle selection was greatly appreciated for added labeling flexibility. E2 mentioned: ''The lasso tool took me a while to get comfortable [with], but now I like it''.

\paragraph{Post-Task Feedback on Usability} 

Table~\ref{tab:nasa_tlx} shows the results of the NASA Task Load Index~\cite{hart2006nasa}, which measures the workload in a range between 0 and 100, with 100 indicating an extremely high workload.
The overall Group Score Results of the NASA Task Load Index was 26.5, suggesting that experts experienced a low-to-moderate cognitive load.
Mental demand scores were the highest, as experts needed to get accustomed to the projection views. 
Ultimately, experts appreciated the data complex relationships the tool encapsulates, as seen by low scores for perceived temporal, effort, and frustration.

Experts gave generally positive feedback on the intuitiveness of \toolname{}. 
When a feature was initially confusing, usage was quickly clarified through experimentation.
The Attribute View and selection percentage details were considered confusing, reflected in E5's frustration score in Table~\ref{tab:nasa_tlx}, highlighting the need for more visual cues.
Experts preferred mouse interactions on the Canvas, rather than control panel pop-ups.
Some minor technical issues negatively impacted the performance score.
Nevertheless, experts reported that \toolname{} enabled them to effectively discover novel patterns which were previously inaccessible to them, as expressed by E5: "I had multiple professional situations where I needed exactly [a tool like] this".

\section{Discussion and Reflection}
\label{sec:discussion_and_reflection}
Reflecting on the process and \toolname{}'s development revealed key discussions on its limitations, design options, generalizability, and future work. 
We offer insights based on our design study methodology, on dataset augmentation via human knowledge externalization, and on the generalizability potential of label-informed projections.

\subsection{User Study Insights}
Participants valued the labeling efficiency gained by their active involvement in the \toolname{} label assignment and validation processes. 
Future studies could involve longer observation periods of experts engaging in comprehensive data labeling, to better understand the challenges and opportunities in data augmentation as the number of expert-labeled particles grows.
How do labeling strategies evolve as experts become more familiar with the tool?
How do experts' workflows evolve when they collaborate on analysis, using \toolname{} as a platform to share their externalized knowledge and generated insights?

\subsection{Scalability}
\toolname{} is the result of an iterative design, with major incremental improvements in client-server communication, rendering algorithms, and particle-rendering speed.
Shifting computation directly to WebGL shaders and using instanced meshes instead of individual geometries in Three.js, increased scalability tremendously.
However, the sheer amount of information in particle images in combination with numerous attributes constitutes a scalability issue, affecting loading times and computational performance. 
Our dataset included 37,857 images of sizes from $10 \times 10$ to $1000 \times 1000$ pixels. 
Informal tests with datasets with 150k images and over double the attributes revealed noticeable performance bottlenecks, as complexity increases linearly with the number of attributes. 
\toolname{}’s reliance on frontend computations, especially with more than 12 attributes, is as a limitation. 
Transitioning some computations to the server side might improve performance.

\subsection{Semi-Supervised Projection vs. Classification} 
We extended interactive data labeling by incorporating semi-supervised dimensionality reduction, allowing labels to influence similarity-preserving particle positioning. 
\toolname{} operates without a classifier, unlike interactive data labeling scenarios that typically pair classifiers with unsupervised dimensionality reduction~\cite{DBLP:journals/tvcg/EirichBJSSFSB22,DBLP:journals/tiis/SevastjanovaJSK21,bernard_jurgen_comparing_2017,Benato2018,pacificVAST2019}. 
However, the value of semi-supervised dimensionality reduction for data labeling is not yet well understood~\cite{visshort2021grossmann}.
Beyond visualizations using unsupervised dimensionality reduction, we see potential of label-informed projections to advocate for more human-model collaboration~\cite{10.1162/neco_a_01434, DBLP:conf/ispa/QianLDCZL21}, leveraging direct feedback through label alphabets. 

\subsection{Feature Ideation through Knowledge Externalization}
\label{sec:discussion:featureIdeation}
Generalizing from our project, not all data attributes/features can always be captured during data abstraction and design already. 
Allowing experts to externalize their knowledge through multiple label alphabets during tool usage provided valuable insights. 
Our method leveraged these label alphabets as additional data attributes/features, integrating the augmented knowledge back into the VA system via label-informed projections. 
We believe this approach holds promise for developing a human-centered methodology for feature ideation~\cite{euroVA2023Schmid}. 
However, the expertise of users is crucial in this process, as their externalized label alphabets will significantly contribute to effective feature ideation.
Our $2\times2$ framework is one way to offer a high degree of control to humans for detecting areas of interest for effective data labeling.

\subsection{Data Labeling: Human Precision or Machine Speed?} 
Our research underscores the crucial balancing the irreplaceable value of human expertise against automation in data labeling.
Automated unsupervised, semi-supervised, and supervised learning methods might overlook the deep insights that experts bring to data analysis. 
Balancing these aspects involves trade-offs: human reliance ensures high accuracy but is time-consuming, while automatic label inference is fast but can introduce inaccuracies. 
\toolname{} enhances human pattern and anomaly discovery by facilitating interactive data analysis and leveraging user-augmented data.
Future developments should thoughtfully weigh the impact on system effectiveness and expert empowerment.

\subsection{Quantification of Tool Success} 
In evaluating \toolname{}, we used validation methods typical for VA and design studies. 
However, long-term benefits remain unassessed, and key metrics such as time savings, cost reductions, and environmental impact remain uncertain.
This reflects a common issue with VA approaches, where the long-term impact of sustainable decision-making and operational efficiency are unclear.
Emphasizing quantifying tool success in real-world applications after extended use could enhance understanding and optimization. 
Notably, Cibulski et al. revisited their PAVED approach after four years of industrial use~\cite{lena2024}, offering valuable insights and paving the way for new validation methodologies.



\section{Conclusion}
We conducted a design study in close collaboration with product quality engineers in medical diagnostics manufacturing, focusing on signs of contamination by different particles.
Understanding their domain practices informed our abstractions and requirements, shaping the design and development of \toolname{}.
Based on a conceptual framework with four main types of expert control, \toolname{} enables contamination analysis by leveraging human labeling as a method to externalize expert knowledge, and using it for insight generation.
\toolname{} facilitates exploratory analysis of thousands of particle data across multiple views, responsive to interactively selected attributes about production details, particle size, and shape characteristics.
Experts can filter and select patterns, label particles with adaptable label alphabets, and augment data with expert knowledge for collaborative decision-making.
Validated in two case studies and a usability user study, \toolname{} has shown potential in improving quality monitoring processes and interactive labeling in medical consumables manufacturing.
We see our approach as making significant progress towards integrating human knowledge into semi-supervised dimensionality reduction, and propose future work in label automation to further study this concept.

\section{Acknowledgments}

We wish to thank our collaborators at Roche Diagnostics Global Operations Consumables, Roche Diagnostics, and Roche Pharma for their invaluable contributions to this work.

\bibliographystyle{abbrv-doi-hyperref}

\bibliography{template}

\begin{thebibliography}{10}

\bibitem{angel2014interactive}
E.~Angel and D.~Shreiner.
\newblock {\em Interactive computer graphics with WebGL}.
\newblock Addison-Wesley Professional, 2014. \href{https://doi.org/10.1145/3532720.3535630}
{doi: {{%
10\hspace{.1pt}\discretionary{.}{%
}{.}\hspace{.4pt}1145\discretionary{/}{%
}{/}3532720\hspace{.1pt}\discretionary{.}{%
}{.}\hspace{.4pt}3535630}}}


\bibitem{attenberg2010}
J.~Attenberg and F.~Provost.
\newblock Inactive learning? {D}ifficulties employing active learning in practice.
\newblock {\em SIGKDD Explor. Newsl.}, 12(2):36–41,  6 pages, 2011. \href{https://doi.org/10.1145/1964897.1964906}
{doi: {{%
10\hspace{.1pt}\discretionary{.}{%
}{.}\hspace{.4pt}1145\discretionary{/}{%
}{/}1964897\hspace{.1pt}\discretionary{.}{%
}{.}\hspace{.4pt}1964906}}}


\bibitem{cagBarth2023}
C.-M. Barth, J.~Schmid, I.~Al-Hazwani, M.~Sachdeva, L.~Cibulski, and J.~Bernard.
\newblock How applicable are attribute-based approaches for human-centered ranking creation?
\newblock {\em Computers \& Graphics (CAG)}, 114:45--58, 2023. \href{https://doi.org/10.1016/j.cag.2023.05.004}
{doi: {{%
10\hspace{.1pt}\discretionary{.}{%
}{.}\hspace{.4pt}1016\discretionary{/}{%
}{/}j\hspace{.1pt}\discretionary{.}{%
}{.}\hspace{.4pt}cag\hspace{.1pt}\discretionary{.}{%
}{.}\hspace{.4pt}2023\hspace{.1pt}\discretionary{.}{%
}{.}\hspace{.4pt}05\hspace{.1pt}\discretionary{.}{%
}{.}\hspace{.4pt}004}}}


\bibitem{Benato2018}
B.~C. Benato, A.~C. Telea, and A.~X. Falcão.
\newblock Semi-supervised learning with interactive label propagation guided by feature space projections.
\newblock In {\em 2018 31st SIBGRAPI Conf. on Graphics, Patterns and Images (SIBGRAPI)}, pp. 392--399, 2018. \href{https://doi.org/10.1109/SIBGRAPI.2018.00057}
{doi: {{%
10\hspace{.1pt}\discretionary{.}{%
}{.}\hspace{.4pt}1109\discretionary{/}{%
}{/}SIBGRAPI\hspace{.1pt}\discretionary{.}{%
}{.}\hspace{.4pt}2018\hspace{.1pt}\discretionary{.}{%
}{.}\hspace{.4pt}00057}}}


\bibitem{vda2017}
J.~Bernard, E.~Dobermann, A.~V{\"o}gele, B.~Kr{\"u}ger, J.~Kohlhammer, and D.~Fellner.
\newblock Visual-interactive semi-supervised labeling of human motion capture data.
\newblock In {\em Visualization and Data Analysis (VDA 2017)}, 2017. \href{https://doi.org/10.2352/ISSN.2470-1173.2017.1.VDA-387}
{doi: {{%
10\hspace{.1pt}\discretionary{.}{%
}{.}\hspace{.4pt}2352\discretionary{/}{%
}{/}ISSN\hspace{.1pt}\discretionary{.}{%
}{.}\hspace{.4pt}2470\discretionary{%
}{-}{-}1173\hspace{.1pt}\discretionary{.}{%
}{.}\hspace{.4pt}2017\hspace{.1pt}\discretionary{.}{%
}{.}\hspace{.4pt}1\hspace{.1pt}\discretionary{.}{%
}{.}\hspace{.4pt}VDA\discretionary{%
}{-}{-}387}}}


\bibitem{euroVisShort2018}
J.~Bernard, M.~Hutter, M.~Lehmann, M.~M\"{u}ller, M.~Zeppelzauer, and M.~Sedlmair.
\newblock {Learning from the Best-Visual Analysis of a Quasi-Optimal Data Labeling Strategy}.
\newblock In {\em EuroVis 2018-Short Papers}. The Eurographics Association, 2018. \href{https://doi.org/10.2312/eurovisshort.20181085}
{doi: {{%
10\hspace{.1pt}\discretionary{.}{%
}{.}\hspace{.4pt}2312\discretionary{/}{%
}{/}eurovisshort\hspace{.1pt}\discretionary{.}{%
}{.}\hspace{.4pt}20181085}}}


\bibitem{bernard_jurgen_comparing_2017}
J.~Bernard, M.~Hutter, M.~Zeppelzauer, D.~W. Fellner, and M.~Sedlmair.
\newblock Comparing visual-interactive labeling with active learning: An experimental study.
\newblock {\em IEEE Trans. Visual Comput. Graphics (TVCG)}, 24(1):298--308, 2018. \href{https://doi.org/10.1109/TVCG.2017.2744818}
{doi: {{%
10\hspace{.1pt}\discretionary{.}{%
}{.}\hspace{.4pt}1109\discretionary{/}{%
}{/}TVCG\hspace{.1pt}\discretionary{.}{%
}{.}\hspace{.4pt}2017\hspace{.1pt}\discretionary{.}{%
}{.}\hspace{.4pt}2744818}}}


\bibitem{ivapp2017Soccer}
J.~Bernard, C.~Ritter, D.~Sessler, M.~Zeppelzauer, J.~Kohlhammer, and D.~Fellner.
\newblock Visual-interactive similarity search for complex objects by example of soccer player analysis.
\newblock In {\em Jt. Conf. Comput. Vis. Imaging Comput. Graph. Theory Appl. (VISGRAP)}, vol.~3, pp. 75--87, 2017. \href{https://doi.org/10.5220/0006116400750087}
{doi: {{%
10\hspace{.1pt}\discretionary{.}{%
}{.}\hspace{.4pt}5220\discretionary{/}{%
}{/}0006116400750087}}}


\bibitem{vahc2015}
J.~Bernard, D.~Sessler, A.~Bannach, T.~May, and J.~Kohlhammer.
\newblock A visual active learning system for the assessment of patient well-being in prostate cancer research.
\newblock In {\em VIS Workshop on Visual Analytics in Healthcare},  article no. 1,  8 pages, pp. 1--8. ACM, 2015. \href{https://doi.org/10.1145/2836034.2836035}
{doi: {{%
10\hspace{.1pt}\discretionary{.}{%
}{.}\hspace{.4pt}1145\discretionary{/}{%
}{/}2836034\hspace{.1pt}\discretionary{.}{%
}{.}\hspace{.4pt}2836035}}}


\bibitem{bernardCGF2018}
J.~Bernard, M.~Zeppelzauer, M.~Lehmann, M.~M\"{u}ller, and M.~Sedlmair.
\newblock Towards user-centered active learning algorithms.
\newblock {\em Comput. Graphics Forum (CGF)}, 37(3):121--132, 2018. \href{https://doi.org/10.1111/cgf.13406}
{doi: {{%
10\hspace{.1pt}\discretionary{.}{%
}{.}\hspace{.4pt}1111\discretionary{/}{%
}{/}cgf\hspace{.1pt}\discretionary{.}{%
}{.}\hspace{.4pt}13406}}}


\bibitem{bernard2018vial}
J.~Bernard, M.~Zeppelzauer, M.~Sedlmair, and W.~Aigner.
\newblock {VIAL}: a unified process for visual interactive labeling.
\newblock {\em Vis. Comput.}, 34(9):1189--1207, 2018. \href{https://doi.org/10.1007/s00371-018-1500-3}
{doi: {{%
10\hspace{.1pt}\discretionary{.}{%
}{.}\hspace{.4pt}1007\discretionary{/}{%
}{/}s00371\discretionary{%
}{-}{-}018\discretionary{%
}{-}{-}1500\discretionary{%
}{-}{-}3}}}


\bibitem{blandford2013semistructured}
A.~Blandford.
\newblock Semi-structured qualitative studies.
\newblock In {\em The Encyclopedia of Human-Computer Interaction, 2nd Ed.} The Interaction Design Foundation, 2013.

\bibitem{blumenschein2018smartexplore}
M.~Blumenschein, M.~Behrisch, S.~Schmid, S.~Butscher, D.~R. Wahl, K.~Villinger, B.~Renner, H.~Reiterer, and D.~A. Keim.
\newblock Smartexplore: Simplifying high-dimensional data analysis through a table-based visual analytics approach.
\newblock In R.~Chang, H.~Qu, and T.~Schreck, eds., {\em IEEE Conf. Visual Analytics Science \& Technology (VAST)}, pp. 36--47. {IEEE}, Berlin, 2018. \href{https://doi.org/10.1109/VAST.2018.8802486}
{doi: {{%
10\hspace{.1pt}\discretionary{.}{%
}{.}\hspace{.4pt}1109\discretionary{/}{%
}{/}VAST\hspace{.1pt}\discretionary{.}{%
}{.}\hspace{.4pt}2018\hspace{.1pt}\discretionary{.}{%
}{.}\hspace{.4pt}8802486}}}


\bibitem{bouali2016vizassist}
F.~Bouali, A.~E. Guettala, and G.~Venturini.
\newblock Vizassist: an interactive user assistant for visual data mining.
\newblock {\em Vis. Comput.}, 32(11):1447--1463, 2016. \href{https://doi.org/10.1007/S00371-015-1132-9}
{doi: {{%
10\hspace{.1pt}\discretionary{.}{%
}{.}\hspace{.4pt}1007\discretionary{/}{%
}{/}S00371\discretionary{%
}{-}{-}015\discretionary{%
}{-}{-}1132\discretionary{%
}{-}{-}9}}}


\bibitem{boukhelifa2013evolutionary}
N.~Boukhelifa, W.~Cancino, A.~Bezerianos, and E.~Lutton.
\newblock Evolutionary visual exploration: evaluation with expert users.
\newblock In {\em Comput. Graphics Forum (CGF)}, vol.~32, pp. 31--40. Wiley Online Library, 2013. \href{https://doi.org/10.1111/cgf.12090}
{doi: {{%
10\hspace{.1pt}\discretionary{.}{%
}{.}\hspace{.4pt}1111\discretionary{/}{%
}{/}cgf\hspace{.1pt}\discretionary{.}{%
}{.}\hspace{.4pt}12090}}}


\bibitem{BuddRK21}
S.~Budd, E.~C. Robinson, and B.~Kainz.
\newblock A survey on active learning and human-in-the-loop deep learning for medical image analysis.
\newblock {\em Medical Image Analysis}, 71:102062, 2021. \href{https://doi.org/10.1016/J.MEDIA.2021.102062}
{doi: {{%
10\hspace{.1pt}\discretionary{.}{%
}{.}\hspace{.4pt}1016\discretionary{/}{%
}{/}J\hspace{.1pt}\discretionary{.}{%
}{.}\hspace{.4pt}MEDIA\hspace{.1pt}\discretionary{.}{%
}{.}\hspace{.4pt}2021\hspace{.1pt}\discretionary{.}{%
}{.}\hspace{.4pt}102062}}}


\bibitem{chang2017revolt}
J.~C. Chang, S.~Amershi, and E.~Kamar.
\newblock Revolt: Collaborative crowdsourcing for labeling machine learning datasets.
\newblock In {\em ACM Conf. Human Factors in Computing Systems (CHI)}, pp. 2334--2346. {ACM}, 2017. \href{https://doi.org/10.1145/3025453.3026044}
{doi: {{%
10\hspace{.1pt}\discretionary{.}{%
}{.}\hspace{.4pt}1145\discretionary{/}{%
}{/}3025453\hspace{.1pt}\discretionary{.}{%
}{.}\hspace{.4pt}3026044}}}


\bibitem{chang2016alloy}
J.~C. Chang, A.~Kittur, and N.~Hahn.
\newblock Alloy: Clustering with crowds and computation.
\newblock In {\em ACM Conf. Human Factors in Computing Systems (CHI)}, pp. 3180--3191. {ACM}, 2016. \href{https://doi.org/10.1145/2858036.2858411}
{doi: {{%
10\hspace{.1pt}\discretionary{.}{%
}{.}\hspace{.4pt}1145\discretionary{/}{%
}{/}2858036\hspace{.1pt}\discretionary{.}{%
}{.}\hspace{.4pt}2858411}}}


\bibitem{Interactive_Lung9394090}
S.~Chatani, Y.~Ma, H.~Zhang, Y.~Chen, and W.~Du.
\newblock Interactive labeling system for lung nodules with ct images.
\newblock In {\em Int. Symp. Comput. Consum. Control}, pp. 529--532, 2020. \href{https://doi.org/10.1109/IS3C50286.2020.00143}
{doi: {{%
10\hspace{.1pt}\discretionary{.}{%
}{.}\hspace{.4pt}1109\discretionary{/}{%
}{/}IS3C50286\hspace{.1pt}\discretionary{.}{%
}{.}\hspace{.4pt}2020\hspace{.1pt}\discretionary{.}{%
}{.}\hspace{.4pt}00143}}}


\bibitem{pacificVAST2019}
M.~Chegini, J.~Bernard, P.~Berger, A.~Sourin, K.~Andrews, and T.~Schreck.
\newblock {Interactive Labelling of a Multivariate Dataset for Supervised Machine Learning Using Linked Visualisations, Clustering, and Active Learning}.
\newblock {\em Visual Informatics}, 3(1):9--17, 2019. \href{https://doi.org/10.1016/j.visinf.2019.03.002}
{doi: {{%
10\hspace{.1pt}\discretionary{.}{%
}{.}\hspace{.4pt}1016\discretionary{/}{%
}{/}j\hspace{.1pt}\discretionary{.}{%
}{.}\hspace{.4pt}visinf\hspace{.1pt}\discretionary{.}{%
}{.}\hspace{.4pt}2019\hspace{.1pt}\discretionary{.}{%
}{.}\hspace{.4pt}03\hspace{.1pt}\discretionary{.}{%
}{.}\hspace{.4pt}002}}}


\bibitem{chipman2000introduction}
S.~F. Chipman, J.~M. Schraagen, and V.~L. Shalin.
\newblock Introduction to cognitive task analysis.
\newblock In {\em Cognitive task analysis}, pp. 17--38. Psychology Press, 2000. \href{https://doi.org/10.4324/9781410605795-9}
{doi: {{%
10\hspace{.1pt}\discretionary{.}{%
}{.}\hspace{.4pt}4324\discretionary{/}{%
}{/}9781410605795\discretionary{%
}{-}{-}9}}}


\bibitem{chung2015knowledge}
D.~H. Chung, M.~L. Parry, I.~W. Griffiths, R.~S. Laramee, R.~Bown, P.~A. Legg, and M.~Chen.
\newblock Knowledge-assisted ranking: A visual analytic application for sports event data.
\newblock {\em IEEE Comput. Graphics Appl.}, 36(3):72--82, 2015. \href{https://doi.org/10.1109/mcg.2015.25}
{doi: {{%
10\hspace{.1pt}\discretionary{.}{%
}{.}\hspace{.4pt}1109\discretionary{/}{%
}{/}mcg\hspace{.1pt}\discretionary{.}{%
}{.}\hspace{.4pt}2015\hspace{.1pt}\discretionary{.}{%
}{.}\hspace{.4pt}25}}}


\bibitem{lena2024}
L.~Cibulski and T.~May.
\newblock {Revisiting PAVED: Studying Tool Adoption After Four Years}.
\newblock In {\em EuroVis-Short Papers}. The Eurographics Association, 2024. \href{https://doi.org/10.2312/evs.20241067}
{doi: {{%
10\hspace{.1pt}\discretionary{.}{%
}{.}\hspace{.4pt}2312\discretionary{/}{%
}{/}evs\hspace{.1pt}\discretionary{.}{%
}{.}\hspace{.4pt}20241067}}}


\bibitem{coscia2024deepsee}
A.~Coscia, H.~M. Sapers, N.~Deutsch, M.~Khurana, J.~S. Magyar, S.~A. Parra, D.~R. Utter, R.~L. Wipfler, D.~W. Caress, E.~J. Martin, J.~B. Paduan, M.~Hendrie, S.~Lombeyda, H.~Mushkin, A.~Endert, S.~Davidoff, and V.~J. Orphan.
\newblock Deepsee: Multidimensional visualizations of seabed ecosystems.
\newblock In {\em ACM Conf. Human Factors in Computing Systems (CHI)}. {ACM}, 2024. \href{https://doi.org/10.1145/3613904.3642001}
{doi: {{%
10\hspace{.1pt}\discretionary{.}{%
}{.}\hspace{.4pt}1145\discretionary{/}{%
}{/}3613904\hspace{.1pt}\discretionary{.}{%
}{.}\hspace{.4pt}3642001}}}


\bibitem{crisanFT21}
A.~Crisan, B.~Fiore{-}Gartland, and M.~Tory.
\newblock Passing the data baton : {A} retrospective analysis on data science work and workers.
\newblock {\em IEEE Trans. Visual Comput. Graphics (TVCG)}, 27(2):1860--1870, 2021. \href{https://doi.org/10.1109/TVCG.2020.3030340}
{doi: {{%
10\hspace{.1pt}\discretionary{.}{%
}{.}\hspace{.4pt}1109\discretionary{/}{%
}{/}TVCG\hspace{.1pt}\discretionary{.}{%
}{.}\hspace{.4pt}2020\hspace{.1pt}\discretionary{.}{%
}{.}\hspace{.4pt}3030340}}}


\bibitem{danchilla2012three}
B.~Danchilla and B.~Danchilla.
\newblock Three. js framework.
\newblock {\em Beginning WebGL for HTML5}, pp. 173--203, 2012.

\bibitem{chartjs}
C.~developers.
\newblock Chart.js.
\newblock \url{github.com/chartjs/Chart.js}, 2015.

\bibitem{Duhaime16}
D.~Duhaime.
\newblock Pixplot.
\newblock \url{github.com/YaleDHLab/pix-plot}, 2016.

\bibitem{DBLP:journals/tvcg/EirichBJSSFSB22}
J.~Eirich, J.~Bonart, D.~J{\"{a}}ckle, M.~Sedlmair, U.~Schmid, K.~Fischbach, T.~Schreck, and J.~Bernard.
\newblock {IRVINE:} {A} design study on analyzing correlation patterns of electrical engines.
\newblock {\em IEEE Trans. Visual Comput. Graphics (TVCG)}, 28(1):11--21, 2022. \href{https://doi.org/10.1109/TVCG.2021.3114797}
{doi: {{%
10\hspace{.1pt}\discretionary{.}{%
}{.}\hspace{.4pt}1109\discretionary{/}{%
}{/}TVCG\hspace{.1pt}\discretionary{.}{%
}{.}\hspace{.4pt}2021\hspace{.1pt}\discretionary{.}{%
}{.}\hspace{.4pt}3114797}}}


\bibitem{endertBN13}
A.~Endert, L.~Bradel, and C.~North.
\newblock Beyond control panels: Direct manipulation for visual analytics.
\newblock {\em IEEE Comput. Graphics Appl. (CGA)}, 33(4):6--13, 2013. \href{https://doi.org/10.1109/MCG.2013.53}
{doi: {{%
10\hspace{.1pt}\discretionary{.}{%
}{.}\hspace{.4pt}1109\discretionary{/}{%
}{/}MCG\hspace{.1pt}\discretionary{.}{%
}{.}\hspace{.4pt}2013\hspace{.1pt}\discretionary{.}{%
}{.}\hspace{.4pt}53}}}


\bibitem{felix2018exploratory}
C.~Felix, A.~Dasgupta, and E.~Bertini.
\newblock The exploratory labeling assistant: Mixed-initiative label curation with large document collections.
\newblock In {\em ACM Symposium on User Interface Software and Technology}, pp. 153--164, 2018. \href{https://doi.org/10.1145/3242587.3242596}
{doi: {{%
10\hspace{.1pt}\discretionary{.}{%
}{.}\hspace{.4pt}1145\discretionary{/}{%
}{/}3242587\hspace{.1pt}\discretionary{.}{%
}{.}\hspace{.4pt}3242596}}}


\bibitem{https://doi.org/10.1111/2041-210X.12584}
O.~Friard and M.~Gamba.
\newblock Boris: a free, versatile open-source event-logging software for video/audio coding and live observations.
\newblock {\em Methods in Ecology and Evolution}, 7(11):1325--1330, 2016. \href{https://doi.org/10.1111/2041-210X.12584}
{doi: {{%
10\hspace{.1pt}\discretionary{.}{%
}{.}\hspace{.4pt}1111\discretionary{/}{%
}{/}2041\discretionary{%
}{-}{-}210X\hspace{.1pt}\discretionary{.}{%
}{.}\hspace{.4pt}12584}}}


\bibitem{DBLP:conf/chi/FruchardMCH23}
B.~Fruchard, S.~Malacria, G.~Casiez, and S.~Huot.
\newblock User preference and performance using tagging and browsing for image labeling.
\newblock In {\em ACM Conf. Human Factors in Computing Systems (CHI)}, pp. 358:1--358:13. {ACM}, 2023. \href{https://doi.org/10.1145/3544548.3580926}
{doi: {{%
10\hspace{.1pt}\discretionary{.}{%
}{.}\hspace{.4pt}1145\discretionary{/}{%
}{/}3544548\hspace{.1pt}\discretionary{.}{%
}{.}\hspace{.4pt}3580926}}}


\bibitem{fujiwara2021interactive}
T.~Fujiwara, X.~Wei, J.~Zhao, and K.-L. Ma.
\newblock Interactive dimensionality reduction for comparative analysis.
\newblock {\em IEEE Trans. Visual Comput. Graphics}, 28(1):758--768, 2021. \href{https://doi.org/10.1109/TVCG.2021.3114807}
{doi: {{%
10\hspace{.1pt}\discretionary{.}{%
}{.}\hspace{.4pt}1109\discretionary{/}{%
}{/}TVCG\hspace{.1pt}\discretionary{.}{%
}{.}\hspace{.4pt}2021\hspace{.1pt}\discretionary{.}{%
}{.}\hspace{.4pt}3114807}}}


\bibitem{DBLP:journals/tvcg/GarrisonMSOHB21}
L.~A. Garrison, J.~M{\"{u}}ller, S.~Schreiber, S.~Oeltze{-}Jafra, H.~Hauser, and S.~Bruckner.
\newblock Dimlift: Interactive hierarchical data exploration through dimensional bundling.
\newblock {\em IEEE Trans. Visual Comput. Graphics (TVCG)}, 27(6):2908--2922, 2021. \href{https://doi.org/10.1109/TVCG.2021.3057519}
{doi: {{%
10\hspace{.1pt}\discretionary{.}{%
}{.}\hspace{.4pt}1109\discretionary{/}{%
}{/}TVCG\hspace{.1pt}\discretionary{.}{%
}{.}\hspace{.4pt}2021\hspace{.1pt}\discretionary{.}{%
}{.}\hspace{.4pt}3057519}}}


\bibitem{visshort2021grossmann}
N.~Grossmann, J.~Bernard, M.~Sedlmair, and M.~Waldner.
\newblock Does the layout really matter? {A} study on visual model accuracy estimation.
\newblock In {\em {IEEE} Vis. Conf.}, pp. 61--65. {IEEE}, 2021. \href{https://doi.org/10.1109/VIS49827.2021.9623326}
{doi: {{%
10\hspace{.1pt}\discretionary{.}{%
}{.}\hspace{.4pt}1109\discretionary{/}{%
}{/}VIS49827\hspace{.1pt}\discretionary{.}{%
}{.}\hspace{.4pt}2021\hspace{.1pt}\discretionary{.}{%
}{.}\hspace{.4pt}9623326}}}


\bibitem{DBLP:journals/mva/HanFNLLN23}
H.~Han, R.~Faust, B.~F.~K. Norambuena, J.~Lin, S.~Li, and C.~North.
\newblock Explainable interactive projections of images.
\newblock {\em Mach. Vis. Appl.}, 34(6):100, 2023. \href{https://doi.org/10.1007/S00138-023-01452-9}
{doi: {{%
10\hspace{.1pt}\discretionary{.}{%
}{.}\hspace{.4pt}1007\discretionary{/}{%
}{/}S00138\discretionary{%
}{-}{-}023\discretionary{%
}{-}{-}01452\discretionary{%
}{-}{-}9}}}


\bibitem{hart2006nasa}
S.~G. Hart.
\newblock Nasa-task load index (nasa-tlx); 20 years later.
\newblock In {\em Proc. Hum. Factors Ergon. Soc. Annu. Meet.}, vol.~50, pp. 904--908. Sage publications Sage CA: Los Angeles, CA, 2006. \href{https://doi.org/10.1037/e577632012-009}
{doi: {{%
10\hspace{.1pt}\discretionary{.}{%
}{.}\hspace{.4pt}1037\discretionary{/}{%
}{/}e577632012\discretionary{%
}{-}{-}009}}}


\bibitem{hoi2006large}
S.~C. Hoi, R.~Jin, and M.~R. Lyu.
\newblock Large-scale text categorization by batch mode active learning.
\newblock In {\em World Wide Web}, pp. 633--642. ACM, 2006. \href{https://doi.org/10.1145/1135777.1135870}
{doi: {{%
10\hspace{.1pt}\discretionary{.}{%
}{.}\hspace{.4pt}1145\discretionary{/}{%
}{/}1135777\hspace{.1pt}\discretionary{.}{%
}{.}\hspace{.4pt}1135870}}}


\bibitem{hoeferlin2012}
B.~Höferlin, R.~Netzel, M.~Höferlin, D.~Weiskopf, and G.~Heidemann.
\newblock Inter-active learning of ad-hoc classifiers for video visual analytics.
\newblock In {\em IEEE Conf. Visual Analytics Science \& Technology (VAST)}, pp. 23--32, 2012. \href{https://doi.org/10.1109/VAST.2012.6400492}
{doi: {{%
10\hspace{.1pt}\discretionary{.}{%
}{.}\hspace{.4pt}1109\discretionary{/}{%
}{/}VAST\hspace{.1pt}\discretionary{.}{%
}{.}\hspace{.4pt}2012\hspace{.1pt}\discretionary{.}{%
}{.}\hspace{.4pt}6400492}}}


\bibitem{park2024dynamiclabels}
P.~Jeongeon, K.~Eun-Young, S.~P. Yeon, Y.~Jinyeong, and K.~Juho.
\newblock Dynamiclabels: Supporting informed construction of machine learning label sets with crowd feedback.
\newblock In {\em {ACM}}, 2024. \href{https://doi.org/10.1145/3640543.3645157}
{doi: {{%
10\hspace{.1pt}\discretionary{.}{%
}{.}\hspace{.4pt}1145\discretionary{/}{%
}{/}3640543\hspace{.1pt}\discretionary{.}{%
}{.}\hspace{.4pt}3645157}}}


\bibitem{keller2021vitessce}
M.~S. Keller, I.~Gold, C.~McCallum, T.~Manz, P.~V. Kharchenko, and N.~Gehlenborg.
\newblock {Vitessce: a framework for integrative visualization of multi-modal and spatially-resolved single-cell data}.
\newblock {\em OSF Preprints}, Oct. 2021. \href{https://doi.org/10.31219/osf.io/y8thv}
{doi: {{%
10\hspace{.1pt}\discretionary{.}{%
}{.}\hspace{.4pt}31219\discretionary{/}{%
}{/}osf\hspace{.1pt}\discretionary{.}{%
}{.}\hspace{.4pt}io\discretionary{/}{%
}{/}y8thv}}}


\bibitem{klein1989critical}
G.~A. Klein, R.~Calderwood, and D.~Macgregor.
\newblock Critical decision method for eliciting knowledge.
\newblock {\em IEEE Trans. Syst. Man Cybern. Part B Cybern.}, 19(3):462--472, 1989. \href{https://doi.org/10.1109/21.31053}
{doi: {{%
10\hspace{.1pt}\discretionary{.}{%
}{.}\hspace{.4pt}1109\discretionary{/}{%
}{/}21\hspace{.1pt}\discretionary{.}{%
}{.}\hspace{.4pt}31053}}}


\bibitem{Kuffel.2021}
A.~Kuffel, A.~Gray, and N.~N. Daeid.
\newblock Impact of metal ions on pcr inhibition and {RT-PCR} efficiency.
\newblock {\em Int. j. leg. med.}, 135(1):63--72, 2021. \href{https://doi.org/10.1007/s00414-020-02363-4}
{doi: {{%
10\hspace{.1pt}\discretionary{.}{%
}{.}\hspace{.4pt}1007\discretionary{/}{%
}{/}s00414\discretionary{%
}{-}{-}020\discretionary{%
}{-}{-}02363\discretionary{%
}{-}{-}4}}}


\bibitem{KumarG20}
P.~Kumar and A.~Gupta.
\newblock Active learning query strategies for classification, regression, and clustering: {A} survey.
\newblock {\em J. Comput. Sci. Technol.}, 35(4):913--945, 2020. \href{https://doi.org/10.1007/S11390-020-9487-4}
{doi: {{%
10\hspace{.1pt}\discretionary{.}{%
}{.}\hspace{.4pt}1007\discretionary{/}{%
}{/}S11390\discretionary{%
}{-}{-}020\discretionary{%
}{-}{-}9487\discretionary{%
}{-}{-}4}}}


\bibitem{lohfink2021knowledge}
A.-P. Lohfink, S.~D.~D. Anton, H.~Leitte, and C.~Garth.
\newblock Knowledge rocks: Adding knowledge assistance to visualization systems.
\newblock {\em IEEE Trans. Visual Comput. Graphics}, 28(1):1117--1127, 2021. \href{https://doi.org/10.1109/tvcg.2021.3114687}
{doi: {{%
10\hspace{.1pt}\discretionary{.}{%
}{.}\hspace{.4pt}1109\discretionary{/}{%
}{/}tvcg\hspace{.1pt}\discretionary{.}{%
}{.}\hspace{.4pt}2021\hspace{.1pt}\discretionary{.}{%
}{.}\hspace{.4pt}3114687}}}


\bibitem{LUGHOFER2012}
E.~Lughofer.
\newblock Hybrid active learning for reducing the annotation effort of operators in classification systems.
\newblock {\em Pattern Recognition}, 45(2):884--896, 2012. \href{https://doi.org/10.1016/j.patcog.2011.08.009}
{doi: {{%
10\hspace{.1pt}\discretionary{.}{%
}{.}\hspace{.4pt}1016\discretionary{/}{%
}{/}j\hspace{.1pt}\discretionary{.}{%
}{.}\hspace{.4pt}patcog\hspace{.1pt}\discretionary{.}{%
}{.}\hspace{.4pt}2011\hspace{.1pt}\discretionary{.}{%
}{.}\hspace{.4pt}08\hspace{.1pt}\discretionary{.}{%
}{.}\hspace{.4pt}009}}}


\bibitem{mcinnes2018umap-software}
L.~McInnes, J.~Healy, N.~Saul, and L.~Grossberger.
\newblock Umap: Uniform manifold approximation and projection.
\newblock {\em The Journal of Open Source Software}, 3(29):861, 2018. \href{https://doi.org/10.21105/joss.00861}
{doi: {{%
10\hspace{.1pt}\discretionary{.}{%
}{.}\hspace{.4pt}21105\discretionary{/}{%
}{/}joss\hspace{.1pt}\discretionary{.}{%
}{.}\hspace{.4pt}00861}}}


\bibitem{militello1998applied}
L.~G. Militello and R.~J. Hutton.
\newblock Applied cognitive task analysis (acta): a practitioner's toolkit for understanding cognitive task demands.
\newblock {\em Ergonomics}, 41(11):1618--1641, 1998. \href{https://doi.org/10.1080/001401398186108}
{doi: {{%
10\hspace{.1pt}\discretionary{.}{%
}{.}\hspace{.4pt}1080\discretionary{/}{%
}{/}001401398186108}}}


\bibitem{mistelbauer2012smart}
G.~Mistelbauer, A.~K{\"o}chl, R.~Schernthaner, I.~Baclija, R.~Schernthaner, S.~Bruckner, M.~Sramek, and M.~E. Gr{\"o}ller.
\newblock Smart super views—a knowledge-assisted interface for medical visualization.
\newblock In {\em 2012 IEEE Conf. on Visual Analytics Science and Technology (VAST)}, pp. 163--172. IEEE, 2012. \href{https://doi.org/10.1109/vast.2012.6400555}
{doi: {{%
10\hspace{.1pt}\discretionary{.}{%
}{.}\hspace{.4pt}1109\discretionary{/}{%
}{/}vast\hspace{.1pt}\discretionary{.}{%
}{.}\hspace{.4pt}2012\hspace{.1pt}\discretionary{.}{%
}{.}\hspace{.4pt}6400555}}}


\bibitem{morgenshtern2023riskfix}
G.~Morgenshtern, A.~Verma, S.~Tonekaboni, R.~Greer, J.~Bernard, M.~Mazwi, A.~Goldenberg, and F.~Chevalier.
\newblock {RiskFix: Supporting Expert Validation of Predictive Timeseries Models in High-Intensity Settings}.
\newblock In {\em EuroVis-Short Papers}. The Eurographics Association, 2023. \href{https://doi.org/10.2312/evs.20231036}
{doi: {{%
10\hspace{.1pt}\discretionary{.}{%
}{.}\hspace{.4pt}2312\discretionary{/}{%
}{/}evs\hspace{.1pt}\discretionary{.}{%
}{.}\hspace{.4pt}20231036}}}


\bibitem{2019_Moritz_et_al}
D.~Moritz, C.~Wang, G.~L. Nelson, H.~Lin, A.~M. Smith, B.~Howe, and J.~Heer.
\newblock Formalizing visualization design knowledge as constraints: Actionable and extensible models in draco.
\newblock {\em IEEE Trans. Visual Comput. Graphics}, 25(1):438--448, 2019. \href{https://doi.org/10.1109/tvcg.2018.2865240}
{doi: {{%
10\hspace{.1pt}\discretionary{.}{%
}{.}\hspace{.4pt}1109\discretionary{/}{%
}{/}tvcg\hspace{.1pt}\discretionary{.}{%
}{.}\hspace{.4pt}2018\hspace{.1pt}\discretionary{.}{%
}{.}\hspace{.4pt}2865240}}}


\bibitem{muhlbacher2013partition}
T.~M{\"u}hlbacher and H.~Piringer.
\newblock A partition-based framework for building and validating regression models.
\newblock {\em IEEE Trans. Visual Comput. Graphics}, 19(12):1962--1971, 2013. \href{https://doi.org/10.1109/tvcg.2013.125}
{doi: {{%
10\hspace{.1pt}\discretionary{.}{%
}{.}\hspace{.4pt}1109\discretionary{/}{%
}{/}tvcg\hspace{.1pt}\discretionary{.}{%
}{.}\hspace{.4pt}2013\hspace{.1pt}\discretionary{.}{%
}{.}\hspace{.4pt}125}}}


\bibitem{DBLP:series/lncs/Munzner08}
T.~Munzner.
\newblock Process and pitfalls in writing information visualization research papers.
\newblock In {\em Information Visualization: Human-Centered Issues and Perspectives}, vol. 4950, pp. 134--153. Springer, 2008. \href{https://doi.org/10.1007/978-3-540-70956-5_6}
{doi: {{%
10\hspace{.1pt}\discretionary{.}{%
}{.}\hspace{.4pt}1007\discretionary{/}{%
}{/}978\discretionary{%
}{-}{-}3\discretionary{%
}{-}{-}540\discretionary{%
}{-}{-}70956\discretionary{%
}{-}{-}5\_6}}}


\bibitem{2020_Nie_et_al}
K.~Nie, P.~Baltzer, B.~Preim, and G.~Mistelbauer.
\newblock {Knowledge-Assisted Comparative Assessment of Breast Cancer using Dynamic Contrast-Enhanced Magnetic Resonance Imaging}.
\newblock {\em Comput. Graphics Forum (CGF)}, 2020. \href{https://doi.org/10.1111/cgf.13959}
{doi: {{%
10\hspace{.1pt}\discretionary{.}{%
}{.}\hspace{.4pt}1111\discretionary{/}{%
}{/}cgf\hspace{.1pt}\discretionary{.}{%
}{.}\hspace{.4pt}13959}}}


\bibitem{DBLP:journals/corr/abs-2305-06809}
T.~Ohm, M.~C. Sol{\`{a}}, A.~Karjus, and M.~Schich.
\newblock Collection space navigator: An interactive visualization interface for multidimensional datasets.
\newblock {\em CoRR}, abs/2305.06809, 2023. \href{https://doi.org/10.48550/arXiv.2305.06809}
{doi: {{%
10\hspace{.1pt}\discretionary{.}{%
}{.}\hspace{.4pt}48550\discretionary{/}{%
}{/}arXiv\hspace{.1pt}\discretionary{.}{%
}{.}\hspace{.4pt}2305\hspace{.1pt}\discretionary{.}{%
}{.}\hspace{.4pt}06809}}}


\bibitem{tissuumap}
N.~Pielawski, A.~Andersson, C.~Avenel, A.~Behanova, E.~Chelebian, A.~Klemm, F.~Nysj{\"o}, L.~Solorzano, and C.~W{\"a}hlby.
\newblock Tissuumaps 3: Improvements in interactive visualization, exploration, and quality assessment of large-scale spatial omics data.
\newblock {\em Heliyon}, 9(5):e15306, 2023. \href{https://doi.org/10.1016/j.heliyon.2023.e15306}
{doi: {{%
10\hspace{.1pt}\discretionary{.}{%
}{.}\hspace{.4pt}1016\discretionary{/}{%
}{/}j\hspace{.1pt}\discretionary{.}{%
}{.}\hspace{.4pt}heliyon\hspace{.1pt}\discretionary{.}{%
}{.}\hspace{.4pt}2023\hspace{.1pt}\discretionary{.}{%
}{.}\hspace{.4pt}e15306}}}


\bibitem{DBLP:conf/ispa/QianLDCZL21}
A.~Qian, C.~Li, X.~Dong, S.~Chen, Y.~Zhang, and G.~Li.
\newblock Slamvis: An interactive visualization approach for smart labeling on multidimensional data.
\newblock In {\em Parallel {\&} Distributed Processing with Applications}, pp. 19--26. {IEEE}, 2021. \href{https://doi.org/10.1109/ISPA-BDCLOUD-SOCIALCOM-SUSTAINCOM52081.2021.00018}
{doi: {{%
10\hspace{.1pt}\discretionary{.}{%
}{.}\hspace{.4pt}1109\discretionary{/}{%
}{/}ISPA\discretionary{%
}{-}{-}BDCLOUD\discretionary{%
}{-}{-}SOCIALCOM\discretionary{%
}{-}{-}SUSTAINCOM52081\hspace{.1pt}\discretionary{.}{%
}{.}\hspace{.4pt}2021\hspace{.1pt}\discretionary{.}{%
}{.}\hspace{.4pt}00018}}}


\bibitem{RAADNUI2005871}
S.~Raadnui.
\newblock Wear particle analysis—utilization of quantitative computer image analysis: A review.
\newblock {\em Tribol. Int.}, 38(10):871--878, 2005. \href{https://doi.org/10.1016/j.triboint.2005.03.013}
{doi: {{%
10\hspace{.1pt}\discretionary{.}{%
}{.}\hspace{.4pt}1016\discretionary{/}{%
}{/}j\hspace{.1pt}\discretionary{.}{%
}{.}\hspace{.4pt}triboint\hspace{.1pt}\discretionary{.}{%
}{.}\hspace{.4pt}2005\hspace{.1pt}\discretionary{.}{%
}{.}\hspace{.4pt}03\hspace{.1pt}\discretionary{.}{%
}{.}\hspace{.4pt}013}}}


\bibitem{Ren22}
P.~Ren, Y.~Xiao, X.~Chang, P.~Huang, Z.~Li, B.~B. Gupta, X.~Chen, and X.~Wang.
\newblock A survey of deep active learning.
\newblock {\em ACM Computing Surveys ({CSUR})}, 54(9):1--40, 2022. \href{https://doi.org/10.1145/3472291}
{doi: {{%
10\hspace{.1pt}\discretionary{.}{%
}{.}\hspace{.4pt}1145\discretionary{/}{%
}{/}3472291}}}


\bibitem{euroVA2018music}
C.~Ritter, C.~Altenhofen, M.~Zeppelzauer, A.~Kuijper, T.~Schreck, and J.~Bernard.
\newblock {Personalized Visual-Interactive Music Classification}.
\newblock In {\em EuroVis Workshop on Visual Analytics (EuroVA)}. The Eurographics Association, 2018. \href{https://doi.org/10.2312/eurova.20181109}
{doi: {{%
10\hspace{.1pt}\discretionary{.}{%
}{.}\hspace{.4pt}2312\discretionary{/}{%
}{/}eurova\hspace{.1pt}\discretionary{.}{%
}{.}\hspace{.4pt}20181109}}}


\bibitem{10.1371/journal.pone.0149856}
U.-P. Rohr, C.~Binder, T.~Dieterle, F.~Giusti, C.~G.~M. Messina, E.~Toerien, H.~Moch, and H.~H. Schäfer.
\newblock The value of in vitro diagnostic testing in medical practice: A status report.
\newblock {\em PLOS ONE}, 11:1--16, 03 2016. \href{https://doi.org/10.1371/journal.pone.0149856}
{doi: {{%
10\hspace{.1pt}\discretionary{.}{%
}{.}\hspace{.4pt}1371\discretionary{/}{%
}{/}journal\hspace{.1pt}\discretionary{.}{%
}{.}\hspace{.4pt}pone\hspace{.1pt}\discretionary{.}{%
}{.}\hspace{.4pt}0149856}}}


\bibitem{cagSachdeva2023}
M.~Sachdeva, J.~Burmeister, J.~Kohlhammer, and J.~Bernard.
\newblock {LFPeers}: Temporal similarity search and result exploration.
\newblock {\em Computers \& Graphics (CAG)}, 2023. \href{https://doi.org/10.1016/j.cag.2023.06.009}
{doi: {{%
10\hspace{.1pt}\discretionary{.}{%
}{.}\hspace{.4pt}1016\discretionary{/}{%
}{/}j\hspace{.1pt}\discretionary{.}{%
}{.}\hspace{.4pt}cag\hspace{.1pt}\discretionary{.}{%
}{.}\hspace{.4pt}2023\hspace{.1pt}\discretionary{.}{%
}{.}\hspace{.4pt}06\hspace{.1pt}\discretionary{.}{%
}{.}\hspace{.4pt}009}}}


\bibitem{10.1162/neco_a_01434}
T.~Sainburg, L.~McInnes, and T.~Q. Gentner.
\newblock {Parametric UMAP Embeddings for Representation and Semisupervised Learning}.
\newblock {\em Neural Computation}, 33(11):2881--2907, 2021. \href{https://doi.org/10.1162/neco_a_01434}
{doi: {{%
10\hspace{.1pt}\discretionary{.}{%
}{.}\hspace{.4pt}1162\discretionary{/}{%
}{/}neco\_a\_01434}}}


\bibitem{euroVA2023Schmid}
J.~Schmidt, H.~Piringer, T.~M{\"u}hlbacher, and J.~Bernard.
\newblock {Human-Based and Automatic Feature Ideation for Time Series Data: A Comparative Study}.
\newblock In {\em EuroVis Workshop on Visual Analytics (EuroVA)}. The Eurographics Association, 2023. \href{https://doi.org/10.2312/eurova.20231089}
{doi: {{%
10\hspace{.1pt}\discretionary{.}{%
}{.}\hspace{.4pt}2312\discretionary{/}{%
}{/}eurova\hspace{.1pt}\discretionary{.}{%
}{.}\hspace{.4pt}20231089}}}


\bibitem{MorphoClusters20113060}
S.-M. Schröder, R.~Kiko, and R.~Koch.
\newblock Morphocluster: Efficient annotation of plankton images by clustering.
\newblock {\em Sensors}, 20(11), 2020. \href{https://doi.org/10.3390/s20113060}
{doi: {{%
10\hspace{.1pt}\discretionary{.}{%
}{.}\hspace{.4pt}3390\discretionary{/}{%
}{/}s20113060}}}


\bibitem{Schwarz1999}
N.~Schwarz.
\newblock Self-reports: How the questions shape the answers.
\newblock {\em American Psychologist}, 54(2):93--105, 1999. \href{https://doi.org/10.1037/0003-066X.54.2.93}
{doi: {{%
10\hspace{.1pt}\discretionary{.}{%
}{.}\hspace{.4pt}1037\discretionary{/}{%
}{/}0003\discretionary{%
}{-}{-}066X\hspace{.1pt}\discretionary{.}{%
}{.}\hspace{.4pt}54\hspace{.1pt}\discretionary{.}{%
}{.}\hspace{.4pt}2\hspace{.1pt}\discretionary{.}{%
}{.}\hspace{.4pt}93}}}


\bibitem{DBLP:conf/beliv/Sedlmair16}
M.~Sedlmair.
\newblock Design study contributions come in different guises: Seven guiding scenarios.
\newblock In {\em Proc. Sixth Workshop Beyond Time Errors Nov. Eval. Methods Vis. (BELIV)}, pp. 152--161. {ACM}, 2016. \href{https://doi.org/10.1145/2993901.2993913}
{doi: {{%
10\hspace{.1pt}\discretionary{.}{%
}{.}\hspace{.4pt}1145\discretionary{/}{%
}{/}2993901\hspace{.1pt}\discretionary{.}{%
}{.}\hspace{.4pt}2993913}}}


\bibitem{DBLP:journals/tvcg/SedlmairMM12}
M.~Sedlmair, M.~D. Meyer, and T.~Munzner.
\newblock Design study methodology: Reflections from the trenches and the stacks.
\newblock {\em IEEE Trans. Visual Comput. Graphics (TVCG)}, 18(12):2431--2440, 2012. \href{https://doi.org/10.1109/TVCG.2012.213}
{doi: {{%
10\hspace{.1pt}\discretionary{.}{%
}{.}\hspace{.4pt}1109\discretionary{/}{%
}{/}TVCG\hspace{.1pt}\discretionary{.}{%
}{.}\hspace{.4pt}2012\hspace{.1pt}\discretionary{.}{%
}{.}\hspace{.4pt}213}}}


\bibitem{seifert2010}
C.~{Seifert} and M.~{Granitzer}.
\newblock User-based active learning.
\newblock In {\em IEEE Int. Conf. on Data Mining Workshops}, pp. 418--425, 2010. \href{https://doi.org/10.1109/ICDMW.2010.181}
{doi: {{%
10\hspace{.1pt}\discretionary{.}{%
}{.}\hspace{.4pt}1109\discretionary{/}{%
}{/}ICDMW\hspace{.1pt}\discretionary{.}{%
}{.}\hspace{.4pt}2010\hspace{.1pt}\discretionary{.}{%
}{.}\hspace{.4pt}181}}}


\bibitem{DBLP:journals/tiis/SevastjanovaJSK21}
R.~Sevastjanova, W.~Jentner, F.~Sperrle, R.~Kehlbeck, J.~Bernard, and M.~El{-}Assady.
\newblock Questioncomb: {A} gamification approach for the visual explanation of linguistic phenomena through interactive labeling.
\newblock {\em {ACM} Trans. Interact. Intell. Syst.}, 11(3-4):19:1--19:38, 2021. \href{https://doi.org/10.1145/3429448}
{doi: {{%
10\hspace{.1pt}\discretionary{.}{%
}{.}\hspace{.4pt}1145\discretionary{/}{%
}{/}3429448}}}


\bibitem{DBLP:journals/tvcg/SomarakisUKLH21}
A.~Somarakis, V.~van Unen, F.~Koning, B.~P.~F. Lelieveldt, and T.~H{\"{o}}llt.
\newblock Imacyte: Visual exploration of cellular micro-environments for imaging mass cytometry data.
\newblock {\em IEEE Trans. Visual Comput. Graphics (TVCG)}, 27(1):98--110, 2021. \href{https://doi.org/10.1109/TVCG.2019.2931299}
{doi: {{%
10\hspace{.1pt}\discretionary{.}{%
}{.}\hspace{.4pt}1109\discretionary{/}{%
}{/}TVCG\hspace{.1pt}\discretionary{.}{%
}{.}\hspace{.4pt}2019\hspace{.1pt}\discretionary{.}{%
}{.}\hspace{.4pt}2931299}}}


\bibitem{2018_Stitz_et_al}
H.~Stitz, S.~Gratzl, H.~Piringer, T.~Zichner, and M.~Streit.
\newblock Knowledgepearls: Provenance-based visualization retrieval.
\newblock {\em IEEE Trans. Visual Comput. Graphics (VAST '18)}, 25(1):120--130, 2018. \href{https://doi.org/10.1109/TVCG.2018.2865024}
{doi: {{%
10\hspace{.1pt}\discretionary{.}{%
}{.}\hspace{.4pt}1109\discretionary{/}{%
}{/}TVCG\hspace{.1pt}\discretionary{.}{%
}{.}\hspace{.4pt}2018\hspace{.1pt}\discretionary{.}{%
}{.}\hspace{.4pt}2865024}}}


\bibitem{wijk05}
J.~J. van Wijk.
\newblock The value of visualization.
\newblock In {\em IEEE Comput.}, pp. 79--86, 2005. \href{https://doi.org/10.1109/VISUAL.2005.1532781}
{doi: {{%
10\hspace{.1pt}\discretionary{.}{%
}{.}\hspace{.4pt}1109\discretionary{/}{%
}{/}VISUAL\hspace{.1pt}\discretionary{.}{%
}{.}\hspace{.4pt}2005\hspace{.1pt}\discretionary{.}{%
}{.}\hspace{.4pt}1532781}}}


\bibitem{WagnerSHRZA19}
M.~Wagner, D.~Slijepcevic, B.~Horsak, A.~Rind, M.~Zeppelzauer, and W.~Aigner.
\newblock Kavagait: Knowledge-assisted visual analytics for clinical gait analysis.
\newblock {\em IEEE Trans. Visual Comput. Graphics (TVCG)}, 25(3):1528--1542, 2019. \href{https://doi.org/10.1109/TVCG.2017.2785271}
{doi: {{%
10\hspace{.1pt}\discretionary{.}{%
}{.}\hspace{.4pt}1109\discretionary{/}{%
}{/}TVCG\hspace{.1pt}\discretionary{.}{%
}{.}\hspace{.4pt}2017\hspace{.1pt}\discretionary{.}{%
}{.}\hspace{.4pt}2785271}}}


\bibitem{wang2024}
H.~Wang, Y.~Ouyang, Y.~Wu, C.~Jiang, L.~Jin, Y.~Cao, and Q.~Li.
\newblock Kmtlabeler: An interactive knowledge-assisted labeling tool for medical text classification.
\newblock {\em IEEE Trans. Visual Comput. Graphics (TVCG)}, pp. 1--18, 2024. \href{https://doi.org/10.1109/TVCG.2024.3406387}
{doi: {{%
10\hspace{.1pt}\discretionary{.}{%
}{.}\hspace{.4pt}1109\discretionary{/}{%
}{/}TVCG\hspace{.1pt}\discretionary{.}{%
}{.}\hspace{.4pt}2024\hspace{.1pt}\discretionary{.}{%
}{.}\hspace{.4pt}3406387}}}


\bibitem{wangJDLRC09}
X.~Wang, D.~H. Jeong, W.~Dou, S.~Lee, W.~Ribarsky, and R.~Chang.
\newblock Defining and applying knowledge conversion processes to a visual analytics system.
\newblock {\em Computers \& Graphics (CAG)}, 33(5):616--623, 2009. \href{https://doi.org/10.1016/j.cag.2009.06.004}
{doi: {{%
10\hspace{.1pt}\discretionary{.}{%
}{.}\hspace{.4pt}1016\discretionary{/}{%
}{/}j\hspace{.1pt}\discretionary{.}{%
}{.}\hspace{.4pt}cag\hspace{.1pt}\discretionary{.}{%
}{.}\hspace{.4pt}2009\hspace{.1pt}\discretionary{.}{%
}{.}\hspace{.4pt}06\hspace{.1pt}\discretionary{.}{%
}{.}\hspace{.4pt}004}}}


\bibitem{DBLP:journals/corr/abs-2306-09328}
Z.~J. Wang, F.~Hohman, and D.~H. Chau.
\newblock Wizmap: Scalable interactive visualization for exploring large machine learning embeddings.
\newblock {\em CoRR}, abs/2306.09328, 2023. \href{https://doi.org/10.48550/arXiv.2306.09328}
{doi: {{%
10\hspace{.1pt}\discretionary{.}{%
}{.}\hspace{.4pt}48550\discretionary{/}{%
}{/}arXiv\hspace{.1pt}\discretionary{.}{%
}{.}\hspace{.4pt}2306\hspace{.1pt}\discretionary{.}{%
}{.}\hspace{.4pt}09328}}}


\bibitem{wenskovitch2021examination}
J.~Wenskovitch and C.~North.
\newblock An examination of grouping and spatial organization tasks for high-dimensional data exploration.
\newblock {\em IEEE Trans. Visual Comput. Graphics (TVCG)}, 27(2):1742--1752, 2021. \href{https://doi.org/10.1109/TVCG.2020.3028890}
{doi: {{%
10\hspace{.1pt}\discretionary{.}{%
}{.}\hspace{.4pt}1109\discretionary{/}{%
}{/}TVCG\hspace{.1pt}\discretionary{.}{%
}{.}\hspace{.4pt}2020\hspace{.1pt}\discretionary{.}{%
}{.}\hspace{.4pt}3028890}}}


\bibitem{sliceProp10403175}
X.~Xu, W.~Lu, J.~Lei, P.~Qiu, H.-B. Shen, and Y.~Yang.
\newblock Sliceprop: A slice-wise bidirectional propagation model for interactive 3d medical image segmentation.
\newblock In {\em Medical Artificial Intelligence (MedAI)}, pp. 414--424, 2023. \href{https://doi.org/10.1109/MedAI59581.2023.00062}
{doi: {{%
10\hspace{.1pt}\discretionary{.}{%
}{.}\hspace{.4pt}1109\discretionary{/}{%
}{/}MedAI59581\hspace{.1pt}\discretionary{.}{%
}{.}\hspace{.4pt}2023\hspace{.1pt}\discretionary{.}{%
}{.}\hspace{.4pt}00062}}}


\bibitem{YangSamuel.2004}
{Yang Samuel} and R.~E. Rothman.
\newblock Pcr-based diagnostics for infectious diseases: uses, limitations, and future applications in acute-care settings.
\newblock {\em The Lancet infectious diseases}, 4:337--348, 06 2004. \href{https://doi.org/10.1016/S1473-3099(04)01044-8}
{doi: {{%
10\hspace{.1pt}\discretionary{.}{%
}{.}\hspace{.4pt}1016\discretionary{/}{%
}{/}S1473\discretionary{%
}{-}{-}3099\discretionary{%
}{(}{(}04\discretionary{)}{%
}{)}01044\discretionary{%
}{-}{-}8}}}


\bibitem{DBLP:journals/tvcg/YuanRWG13}
X.~Yuan, D.~Ren, Z.~Wang, and C.~Guo.
\newblock Dimension projection matrix/tree: Interactive subspace visual exploration and analysis of high dimensional data.
\newblock {\em IEEE Trans. Visual Comput. Graphics (TVCG)}, 19(12):2625--2633, 2013. \href{https://doi.org/10.1109/TVCG.2013.150}
{doi: {{%
10\hspace{.1pt}\discretionary{.}{%
}{.}\hspace{.4pt}1109\discretionary{/}{%
}{/}TVCG\hspace{.1pt}\discretionary{.}{%
}{.}\hspace{.4pt}2013\hspace{.1pt}\discretionary{.}{%
}{.}\hspace{.4pt}150}}}


\bibitem{Zhang2022}
Z.~Zhang, E.~Strubell, and E.~H. Hovy.
\newblock A survey of active learning for natural language processing.
\newblock {\em CoRR}, abs/2210.10109, 2022. \href{https://doi.org/10.48550/ARXIV.2210.10109}
{doi: {{%
10\hspace{.1pt}\discretionary{.}{%
}{.}\hspace{.4pt}48550\discretionary{/}{%
}{/}ARXIV\hspace{.1pt}\discretionary{.}{%
}{.}\hspace{.4pt}2210\hspace{.1pt}\discretionary{.}{%
}{.}\hspace{.4pt}10109}}}


\bibitem{DBLP:journals/tac/ZhuKKST18}
L.~Zhu, P.~Karasev, I.~Kolesov, R.~Sandhu, and A.~R. Tannenbaum.
\newblock Guiding image segmentation on the fly: Interactive segmentation from a feedback control perspective.
\newblock {\em {IEEE} Trans. Autom. Control.}, 63(10):3276--3289, 2018. \href{https://doi.org/10.1109/TAC.2018.2792328}
{doi: {{%
10\hspace{.1pt}\discretionary{.}{%
}{.}\hspace{.4pt}1109\discretionary{/}{%
}{/}TAC\hspace{.1pt}\discretionary{.}{%
}{.}\hspace{.4pt}2018\hspace{.1pt}\discretionary{.}{%
}{.}\hspace{.4pt}2792328}}}


\bibitem{DBLP:journals/tmi/ZhuCMP03}
Y.~Zhu, B.~Carragher, F.~Mouche, and C.~S. Potter.
\newblock Automatic particle detection through efficient hough transforms.
\newblock {\em {IEEE} Trans. Medical Imaging}, 22(9):1053--1062, 2003. \href{https://doi.org/10.1109/TMI.2003.816947}
{doi: {{%
10\hspace{.1pt}\discretionary{.}{%
}{.}\hspace{.4pt}1109\discretionary{/}{%
}{/}TMI\hspace{.1pt}\discretionary{.}{%
}{.}\hspace{.4pt}2003\hspace{.1pt}\discretionary{.}{%
}{.}\hspace{.4pt}816947}}}


\end{thebibliography}

\appendix 

\end{document}